\documentclass[a4paper,11pt]{article}
\pdfoutput=1
\usepackage{jheppub}
\usepackage{subfigure}
\usepackage{relsize}

\def  \bcen   {\begin{center}}
\def  \ecen   {\end{center}}
\def  \beq    {\begin{equation}}
\def  \eeq    {\end{equation}}
\def  \beqa   {\begin{eqnarray}}
\def  \eeqa   {\end{eqnarray}}

\def\bea{\begin{eqnarray}}
\def\eea{\end{eqnarray}}

\title{Influence of anomalous VVH and VVHH on determination of Higgs self couplings at ILC}
\author[1]{Satendra Kumar}
\author[2]{P. Poulose}

\affiliation{Indian Institute of Technology Guwahati, Assam 781039, INDIA}

\emailAdd{satendra@iitg.ernet.in}
\emailAdd{poulose@iitg.ernet.in}
\abstract{ 
The recent discovery of a Higgs boson at LHC, while establishing the Higgs mechanism as the way  of electroweak symmetry breaking, started an era of precision measurements  involving the Higgs boson. In an effective Lagrangian framework, we consider the $e^+e^-\rightarrow ZHH$ process, at an ILC running at a centre of mass energy of 800 GeV to investigate the effect of the $ZZH$ and $ZZHH$ couplings on the sensitivity of $HHH$ coupling on this process. Our results show that  the sensitivity of the trilinear Higgs self couplings on this process has somewhat strong dependence on the Higgs-gauge boson couplings. Single and two parameter reach of ILC with integrated luminosity of 1000 fb$^{-1}$ are obtained on the effective couplings, $c_6$ and $c_H$, which are related to the $HHH$ couplings, indicating how these limits are affected by the presence of anomalous $ZZH$ and $ZZHH$ couplings. The kinematic distributions studied to understand the effect of the anomalous couplings, again, show strong influence of $Z$-$H$ couplings on the dependence of these distributions on $HHH$ coupling. Similar results are indicated in the case of the process, $e^+e^-\rightarrow \nu\bar \nu HH$, considered at a centre of mass energy of 2 TeV, where the cross section is large enough.  The effect of $WWH$ and $WWHH$ couplings on the sensitivity of $HHH$ coupling is clearely established through our analyses of this process.
}


\begin{document}
\maketitle
\section{Introduction}
With the discovery of the new resonance of mass around 125 GeV at LHC \cite{cms, atlas, Moriond1, Moriond2, Moriond3, Moriond4, Moriond5,Aad:2013wqa,Chatrchyan:2012ufa,Chatrchyan:2013lba,Aad:2012tfa}, a new era is open in the investigations of elementary particle dynamics. The new particle is so far consistent in every way with the long expected Higgs boson of the Standard Model (SM). All the expected SM decays are observed at LHC, albeit some tension in the decay widths, which are still consistent within the statistical fluctuations. The spin and parity analysis favour a spin-zero, even-parity object \cite{Moriond1,Moriond2, Moriond3, Moriond4, Moriond5}. Thus, it is perhaps correct to state that the newly observed state is indeed a Higgs boson, establishing that the weakly interactly Higgs mechanism, if not entirely responsible, has a major role in the electroweak symmetry breaking (EWSB). While we wait for further statistics to establish more detailed identity of this excitement, it is worth revisiting the role of new physics in the Higgs sector in the light of the new measurements. It is well accepted that, even if all the properties of the new particle meets the expectations of the SM, there still remain several questions on the SM. One of the serious issues within the Higgs sector is the difficulty with quadratically diverging quantum corrections to the mass of the Higgs boson, or the so called hierarchy problem. This itself should convince us that the SM is at the most an effective theory, highly successfull at the electroweak scale. Among the plethora of suggestions to look beyond the SM, one could indeed narrow down to scenarios that can accommodate a light Higgs boson, very likely an elemenraty one, with properties very close to that of the SM Higgs boson. One may need to wait till LHC reveals further indications of new physics, if we are lucky, or perhaps even need to wait till the new generation lepton colliders, like the International Linear Collider (ILC) \cite{ILC1, ILC2,polarizationreview}, start exploring the TeV scale physics. Being a discovery machine, LHC is capable of observing any direct production of new particle resonance at the energy scales explored, while the latter is more suitable to explore the new physics through detailed precision analysis, in the absence of any such direct observation of new physics. 

Taking cue from the observations so far, one is somewhat compelled to consider a case with new physics somewhat decoupled from the electroweak physics, which in turn is dictated by the SM. In that case, the effect of new physics will be reflected in the various couplings through the quantum corrections they acquire. The best way to study such effects is through an effective Lagrangian, which encodes the new physics effects in higher dimensional operators with anomalous couplings. Interesting phenomenological studies with effective Higgs couplings, including the possibility of CP violation in the Higgs sector is discussed in the literature \footnote{An important issue of the top quark Yukawa coupling in the context of CP-mixed Higgs boson is studies in Ref. \cite{Ananthanarayan:2013cia, Ananthanarayan:2014eea,Muhlleitner:2012jy,Godbole:2011hw}}. 
The study of Higgs sector through an effective Lagrangian, and effective couplings goes back to Refs.\cite{Weinberg:1978kz,Weinberg:1980wa,Georgi:1994qn,Buchmuller:1985jz,Hagiwara:1993ck,Hagiwara:1993qt,Alam:1997nk,
genuined6,Giudice:2007fh,Contino2010a,Contino2010b,Grober2011,Grzadkowski:2010es,
GutierrezRodriguez:2011gi,GutierrezRodriguez:2009uz,GutierrezRodriguez:2005fe,Rindani:2010pi,Rindani:2009pb}.
 More recently,  the Lagrangian including complete set of dimension-6 operators is studied by 
Refs. \cite{Baak:2012kk, Einhorn:2013kja,Contino:2013kra,Amar:2014fpa,Masso:2014xra,Biekoetter:2014jwa,Willenbrock:2014bja}. For some of the recent reference discussing the constraints on the anomalous couplings within different approches, please see 
\cite{Bonnet:2011yx,Corbett:2012dm,Chang:2013cia,Elias-Miro:2013mua,Banerjee:2013apa,Boos:2013mqa,
Masso:2012eq,Han:2004az, Corbett:2012ja,Dumont:2013wma, Pomarol:2013zra,Ellis:2014dva, Belusca-Maito:2014dpa, Gupta:2014rxa}.
Ref. \cite{Ellis:2014dva} studied the H+V, where V= Z, W, associated production at LHC and TeVatron to discuss the bounds obtainable from the global fit to the presently available data, whereas Ref.~\cite{Belusca-Maito:2014dpa} has discussed the constraint on the parameters coming from LHC results as well as other precision data from LEP, SLC and TeVatron. 
Experimental studies on the Higgs couplings at LHC are presented in, for example, \cite{ Aad:2013wqa,Teyssier:2014hta}.

Higgs self couplings give direct information about the scalar potential, and therefore, very important to understand the nature of the EWSB. The process, $e^+ e^-\rightarrow ZHH$ is one of the best suited to study the Higgs  trilinear coupling \cite{De Rujula:1991se,1,2,3,4,5,6,7,8,9,10}.  At the same time, this process also depends on the Higgs-Gauge boson couplings, $ZZH$ and $ZZHH$, which will affect the determination of the the $HHH$ coupling. Another process that could probe the $HHH$ couplings is $e^+e^- \rightarrow \nu \bar{\nu} HH$ following the WW fusion \cite{4,5,6,7}, which is also affected by the $WWH$ and $WWHH$ couplings. In this report we will focus our attention on these processes in some detail within the framework of the effective Lagrangian. One goal of this study is to investigate how significant is the effect of $VVH$ coupling, where $V=Z, ~W$, in the extraction of  the $HHH$ coupling .

The report is presented in the following way. In Section~\ref{sec:setup} the effective Lagrangian will be presented, with the currently available constraint on the parameters. In Section~\ref{sec:discussions} the processes under consideration will be presented, with details. In Section~\ref{sec:summary} the results will be summarized. 

\section{General Setup}\label{sec:setup}
The effective Lagrangian with full set of dimension-6 operator involving the Higgs bosons is described in Ref. ~\cite{Giudice:2007fh,Contino2010a,Contino2010b,Grober2011,Contino:2013kra,Ellis:2014dva}. In this report we shall restrict our discussion to the processes $e^+e^- \rightarrow ZHH$, and $e^+e^- \rightarrow \nu\bar{\nu}WW\rightarrow \nu\bar{\nu}HH$. Relevant to these processes, part of the Lagrangian is given by
\begin{eqnarray}
{\cal L}_{\rm Higgs}^{\rm anom} &=&    \frac{\bar c_{H}}{2 v^2} \partial^\mu\big(\Phi^\dag \Phi\big) \partial_\mu \big( \Phi^\dagger \Phi \big) +  \frac{\bar c_6}{v^2}\lambda~\big(\Phi^\dag\Phi\big)^3+\frac{\bar c_{\gamma}}{m_W^2} g'^2~\Phi^\dag \Phi B_{\mu\nu} B^{\mu\nu}
   + \frac{\bar  c_{g}}{m_W^2}g_s^2 ~\Phi^\dag \Phi G_{\mu\nu}^a G_a^{\mu\nu} \nonumber\\
  &&+ \frac{\bar c_{HW}}{m_W^2} ig~\big(D^\mu \Phi^\dag \sigma_{k} D^\nu \Phi\big) W_{\mu \nu}^k 
  + \frac{\bar c_{HB}}{m_W^2} ig'~ \big(D^\mu \Phi^\dag D^\nu \Phi\big) B_{\mu \nu} \nonumber\ \\
  && + \frac{\bar c_{W}}{2m_W^2} ig~\big( \Phi^\dag \sigma_{k} \overleftrightarrow{D}^\mu \Phi \big)  D^\nu  W_{\mu \nu}^k 
  +\frac{\bar c_{B}}{2 m_W^2} ig'~\big(\Phi^\dag \overleftrightarrow{D}^\mu \Phi \big) \partial^\nu  B_{\mu \nu},
\label{eq:Leff}
\end{eqnarray}
where
\(
  \Phi^\dag {\overleftrightarrow D}_\mu \Phi = 
    \Phi^\dag D^\mu \Phi - D_\mu\Phi^\dag \Phi \ ,
\) $D^\mu$ being the appropriate covarient derivative operator, and $\Phi$, the usual Higgs doublet in the SM. Also, $G_{\mu\nu}^a$, $W_{\mu\nu}^k$ and $B_{\mu\nu}$ are the field tensors corresponding to the $SU(3)_C$, $SU(2)_L$ and $U(1)_Y$ of the SM gauge groups, respectively, with gauge coplings $g_s$, $g$ and $g'$, in that order. $\sigma_k$ are the Pauli matrices, and $\lambda$ is the usual (SM) quadratic coupling constant of the Higgs field.
The above Lagrangian, leads to the following in the unitary gauge and mass basis
~\cite{HEL}
\begin{eqnarray}
  {\cal L}_{H,Z,W} ^{\rm anom}=&& -v \lambda g_{HHH}^{(1)} H^3 + \frac{1}{2} g_{HHH}^{(2)} H \partial_{\mu} H \partial^{\mu} H -\frac{1}{4} g_{HZZ}^{(1)} Z_{\mu\nu} Z^{\mu\nu} H-
 \frac{1}{4} g_{HZZ}^{(2)} Z_{\nu}\partial_{\mu} Z^{\mu\nu} H \nonumber\\
 &&+\frac{1}{2} g_{HZZ}^{(3)} Z_{\mu} Z^{\mu} H -\frac{1}{2} g_{HAZ}^{(1)} Z_{\mu\nu} F^{\mu\nu} H-g_{HAZ}^{(2)}Z_{\nu} \partial_{\mu} F^{\mu\nu} H \nonumber\\ 
 && - \frac{1}{8} g_{HHZZ}^{(1)} Z_{\mu\nu} Z^{\mu\nu} H^2 - \frac{1}{2} g_{HHZZ}^{(2)}Z_{\nu} \partial_{\mu} Z^{\mu\nu} H^2 - \frac{1}{4} g_{HHZZ}^{(3)} Z_{\mu} Z^{\mu} H^2 \nonumber\\
 && - \frac{1}{2} g_{HWW}^{(1)} W^{\mu\nu} W_{\mu\nu}^{\dagger} H -\left [g_{HWW}^{(2)} W^{\nu} \partial^{\mu} W_{\mu\nu}^{\dagger} H + h.c.\right] + g~m_{W} W_{\mu}^{\dagger} W^{\mu} H \nonumber\\ 
 && - \frac{1}{4} g_{HHWW}^{(1)} W^{\mu\nu} W_{\mu\nu}^{\dagger} H^2 - \frac{1}{2} \left[g_{HHWW}^{(2)} W^{\nu} \partial^{\mu} W_{\mu\nu}^{\dagger} H^2 + h.c.\right] + \frac{1}{4} g^2 W_{\mu}^{\dagger} W^{\mu} H^2 
 \nonumber\\
\label{eq:LagPhys}
\end{eqnarray}

Various physical couplings present in the Lagrangian in Eq.~\ref{eq:LagPhys} are given in terms of the parameters of the effective Lagrangian in Eq.~\ref{eq:Leff} as

\begin{eqnarray}
&& g_{HHH}^{(1)} = 1+ \frac{5}{2} \bar{c_6},\hspace{1cm} g_{HHH}^{(2)} = \frac{g}{m_W} \bar{c}_H \nonumber\\
&& g_{HZZ}^{(1)}  = \frac{2g}{c_W^2 m_W} \left[ \bar{c}_{HB} s_W^2 - 4\bar{c}_{\gamma} s_W^4 + c_W^2 \bar{c}_{HW}  \right]\nonumber\\  
&& g_{HZZ}^{(2)}  = \frac{g}{c_W^2 m_W} \left[ (\bar{c}_{HW}+ \bar{c}_W) c_W^2 + (\bar{c}_B + \bar{c}_{HB}) s_W^2  \right],~~~~~g_{HZZ}^{(3)}  = \frac{g m_Z}{c_W} \left[ 1-2 \bar{c}_T  \right] \nonumber\\
&& g_{HAZ}^{(1)}  = \frac{g s_W}{c_W m_W} \left[ \bar{c}_{HW} -\bar{c}_{HB} + 8\bar{c}_{\gamma} s_W^2   \right]\nonumber\\ 
&& g_{HAZ}^{(2)}  = \frac{g s_W}{c_W m_W} \left[ \bar{c}_{HW} -\bar{c}_{HB} -\bar{c}_B + \bar{c}_W   \right] \nonumber\\
&&g_{HHZZ}^{(1)}  = \frac{g^2}{c_W^2 m_W^2} \left[ \bar{c}_{HB} s_W^2 - 4\bar{c}_{\gamma} s_W^4 + \bar{c}_{HW} c_W^2   \right]\nonumber\\
&&g_{HHZZ}^{(2)}  = \frac{g^2}{2 c_W^2 m_W^2} \left[ (\bar{c}_{HW} + \bar{c}_{W}) c_W^2 + (\bar{c}_{B} + \bar{c}_{HB}) s_W^2  \right],~~~~~   g_{HHZZ}^{(3)}  = \frac{g^2}{2 c_W^2} \left[1 - 6 \bar{c}_{T} \right]\nonumber\\
&&g_{HWW}^{(1)}  = \frac{2g}{m_W} \bar{c}_{HW}, ~~~~~~~~ g_{HWW}^{(2)}  = \frac{g}{2m_W} \left[\bar{c}_{W} + \bar{c}_{HW}\right] \nonumber\\
&&g_{HHWW}^{(1)}  = \frac{g^2}{m_W^2} \bar{c}_{HW}, ~~~~~~ g_{HHWW}^{(2)}  = \frac{g^2}{4 m_W^2} \left[\bar{c}_{W} + \bar{c}_{HW}\right] 
\end{eqnarray}

In total eight coefficients, namely, $\bar{c}_6, ~\bar{c}_H,~ \bar{c}_T,~ \bar{c}_{\gamma},~ \bar{c}_B, ~\bar{c}_W,~ \bar{c}_{HB},~ \bar{c}_{HW} $, govern the dyanmics of $ZHH$ and $\nu\bar\nu HH$ productions at ILC. Coming to the experimental constraints on these parameters, the first two, $\bar{c}_6$ and $\bar{c}_H$ influence only the Higgs self couplings, and therefore, practically, do not have any experimental constraints on them. 
Electroweak precision tests constrain $\bar c_T$, $\bar c_W$ and $\bar c_B$ as ~\cite{Baak:2012kk}

\begin{eqnarray}
 &&\bar{c}_T(m_Z) \in [-1.5, 2.2] \times 10^{-3}, \nonumber \\
&&(\bar{c}_W(m_Z) + \bar{c}_B(m_Z))\in [-1.4, 1.9]\times 10^{-3}.
\end{eqnarray}

Note that,  $\bar{c}_W$ and $\bar{c}_B$ are not independently constrained, leaving possibility of having  large values with cancellation between them as per the above constraint. $\bar c_W$ itself, along with $\bar c_{HW}$ and $\bar c_{HB}$ is constrained from LHC observations on associated production of Higgs along with W in Ref. ~\cite{Ellis:2014dva}.
Consideration of the Higgs associated production along with W, ATLAS and CMS along with D0 put a limit of  \( \bar{c}_W \in \big[-0.05, 0.04\big] \), when all other parameters are set to zero. A global fit using various information from ATLAS and CMS, including signal-srength information constrains the region in $\bar c_W-\bar c_{HW}$ plane, leading to a slightly more relaxed limit on $\bar c_W$, and a limit of about \( \bar{c}_{HW} \in \big[-0.1, 0.06\big] \).  The limit on $\bar{c}_{HB}$ estimated using a global fit in Ref.~\cite{Ellis:2014dva} is about \(\bar{c}_{HB} \in [-0.05, 0.05]\) with a one parameter fit.

The purpose of this study is to understand how to exploit a precision machine like the ILC to investigate suitable processes so as to derive information regading these couplings. In the next section we shall explain the processes of interest in the present case, and discuss the details to undestand the influence of one or more of the couplings mentioned above.

\section{Discussion of the processes considered}\label{sec:discussions}

\begin{figure}[h]
\begin{center}
\begin{tabular}{c c}
\includegraphics[width=50mm]{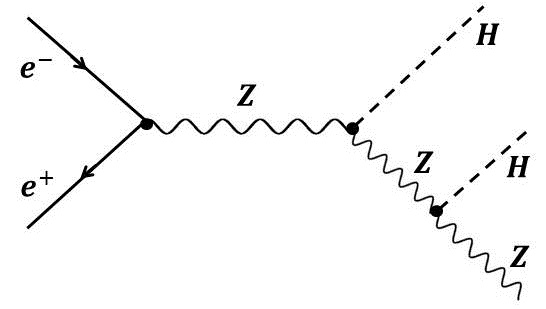}&
\hspace{18mm}
\includegraphics[width=50mm]{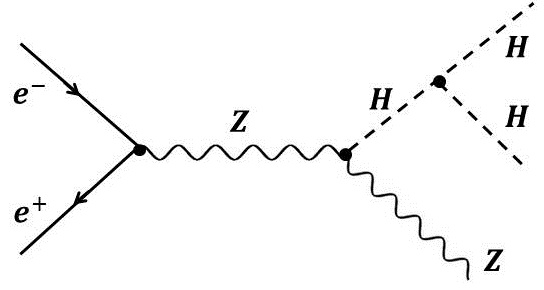}\\[5mm]
\includegraphics[width=50mm]{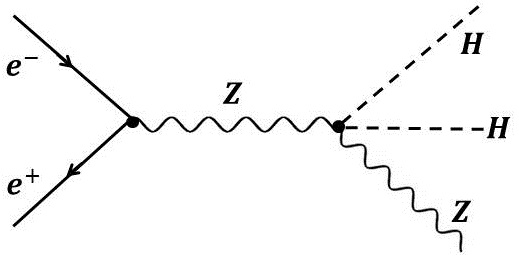}
\hspace{-80mm}
\end{tabular}
\end{center}
\caption{Feynman diagrams contributing to  the process $e^{-}e^+ \rightarrow ZHH$ in Standard Model. }
\label{fig:fdzhh}
\end{figure}

It is generally expected that the ILC, with its clean environment, fixed centre of mass energy, and additional features like availability of beam polarization, will be able to do the precision studies much more efficiently than what LHC could do.  This is especially so in the case of Higgs self couplings. One of the best suited process to study the trilinear (self) coupling of the Higgs boson is $e^+e^- \rightarrow ZHH$, the phenomenological analysis of which is studied in detail within the context of the SM. The Feynman diagrams corresponding to this process in the SM are given in  Fig.~\ref{fig:fdzhh}.

\begin{figure}[h]
\begin{center}
\begin{tabular}{c c}
\includegraphics[width=53mm]{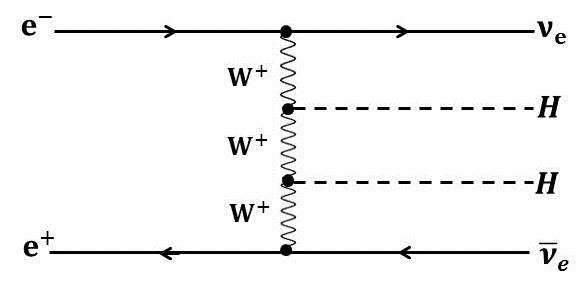}&
\hspace{15mm}
\includegraphics[angle=0,width=53mm]{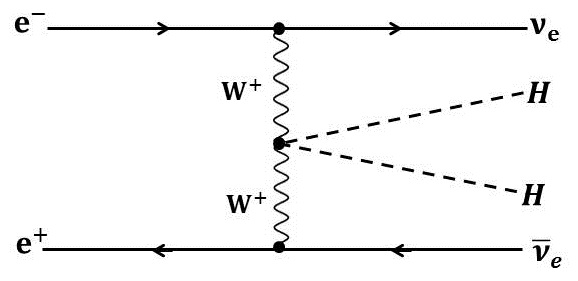} \\[5mm] 
\includegraphics[width=53mm]{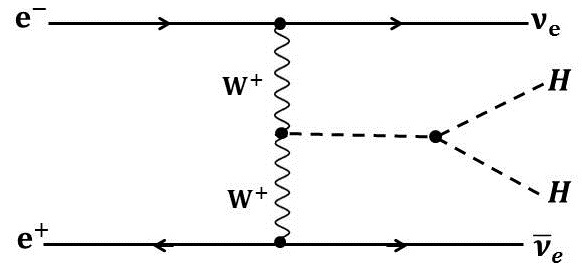}
\hspace{-80mm}
\end{tabular}
\end{center}
\caption{Feynman diagrams contributing to  the process $e^{-}e^+ \rightarrow  \nu\bar\nu HH$ in Standard Model, without considering $e^+e^-\rightarrow ZHH \rightarrow \nu\bar\nu HH $}
\label{fig:fdnnhh}
\end{figure}

Another process that is relevant to the study of $HHH$ coupling is $e^+e^- \rightarrow \nu_e \bar \nu_e HH$. The earlier process, $e^+e^- \rightarrow ZHH$, with the invisible decay of $Z\rightarrow \nu_e\bar\nu_e$ also leads to the same final state. However, this can be easily reduced by considering the missing invariant mass. The rest of the process goes through the Feynman diagrams presented in Fig.~\ref{fig:fdnnhh}.

Apart from the $HHH$ coupling, these processes are influenced by gauge-Higgs couplings like $ZZH$, $ZZHH$, $WWH$ and $WWHH$. Keeping in mind the above discussion of the effective couplings deviating from the SM due to the influence of the BSM at some higher energies, one must understand how such a scenario would affect the phenomenology, in order to draw any conclusion regarding these couplings. In the rest of this report we shall revisit these processes, with a specific purpose of understanding the correlation between the  gauge-Higgs coupling and the trilinear Higgs couplings. 

For our analyses we use MADGRAPH \cite{madgraph}, with the Effective Lagrangian implemented through Feynfules \cite{feynrules} as given by \cite{HEL}.

\subsection{$e^+e^-\rightarrow ZHH$ Process}

\begin{figure}[h] \centering
\begin{tabular}{c c}
\hspace{-5mm}
\includegraphics[angle=0,width=100mm]{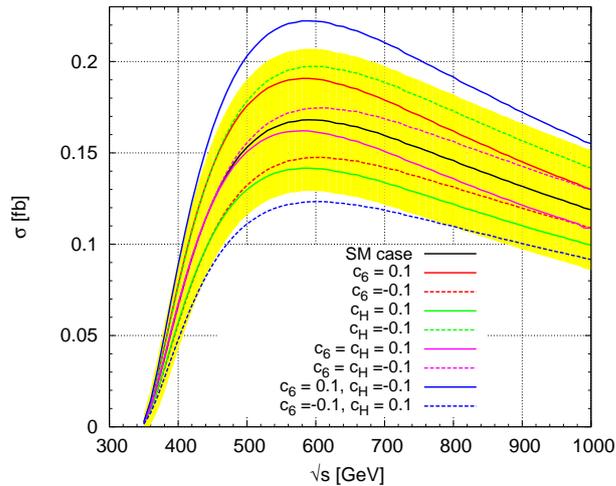} 
\end{tabular}
\caption{Cross section against $\sqrt{s}$ for  the process $e^{-}e^+ \rightarrow ZHH$, for different values of the parameters $c_6$ and $c_H$, with all others kept to zero. }
\label{fig:cs_roots_zhh}
\end{figure}

We shall first consider $e^+e^- \rightarrow ZHH$ process. In Fig.~\ref{fig:cs_roots_zhh} the cross section is plotted against the centre of mass for the SM case as well as for some selected $(c_6, c_H)$ points. The cross section peaks around a centre of mass energy of 600 GeV with  a value of about 0.17 fb, which slides down to about 0.15 fb at 800 GeV. In order to avoid any complications arising from the threshold effects, we perform our analysis for an ILC running at a centre of mass energy of 800 GeV, sufficiently away from the threshold value.  

\begin{figure}[h]\centering
\begin{tabular}{c c}
\hspace{-5mm}
\includegraphics[angle=0,width=80mm]{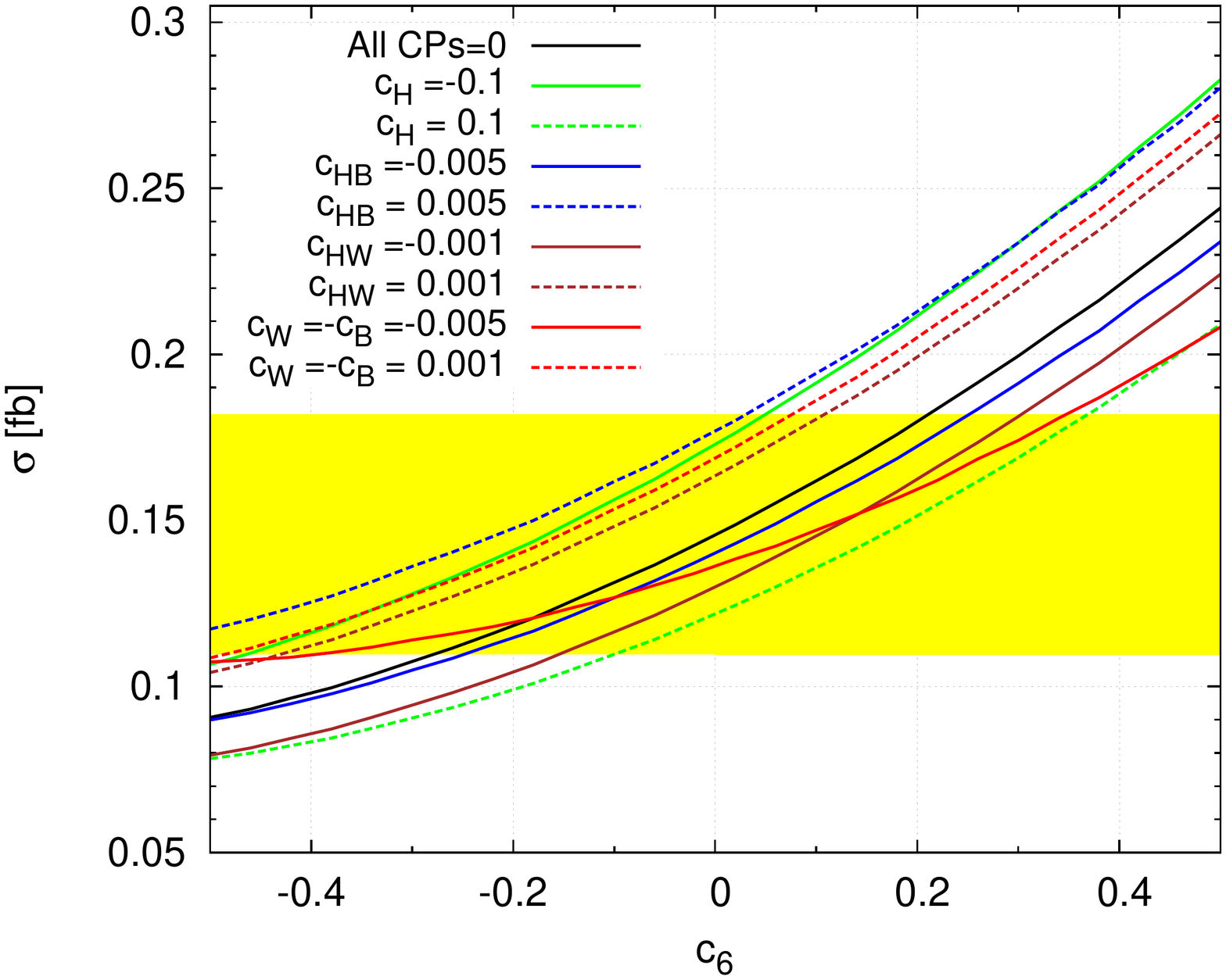}&
\includegraphics[angle=0,width=80mm]{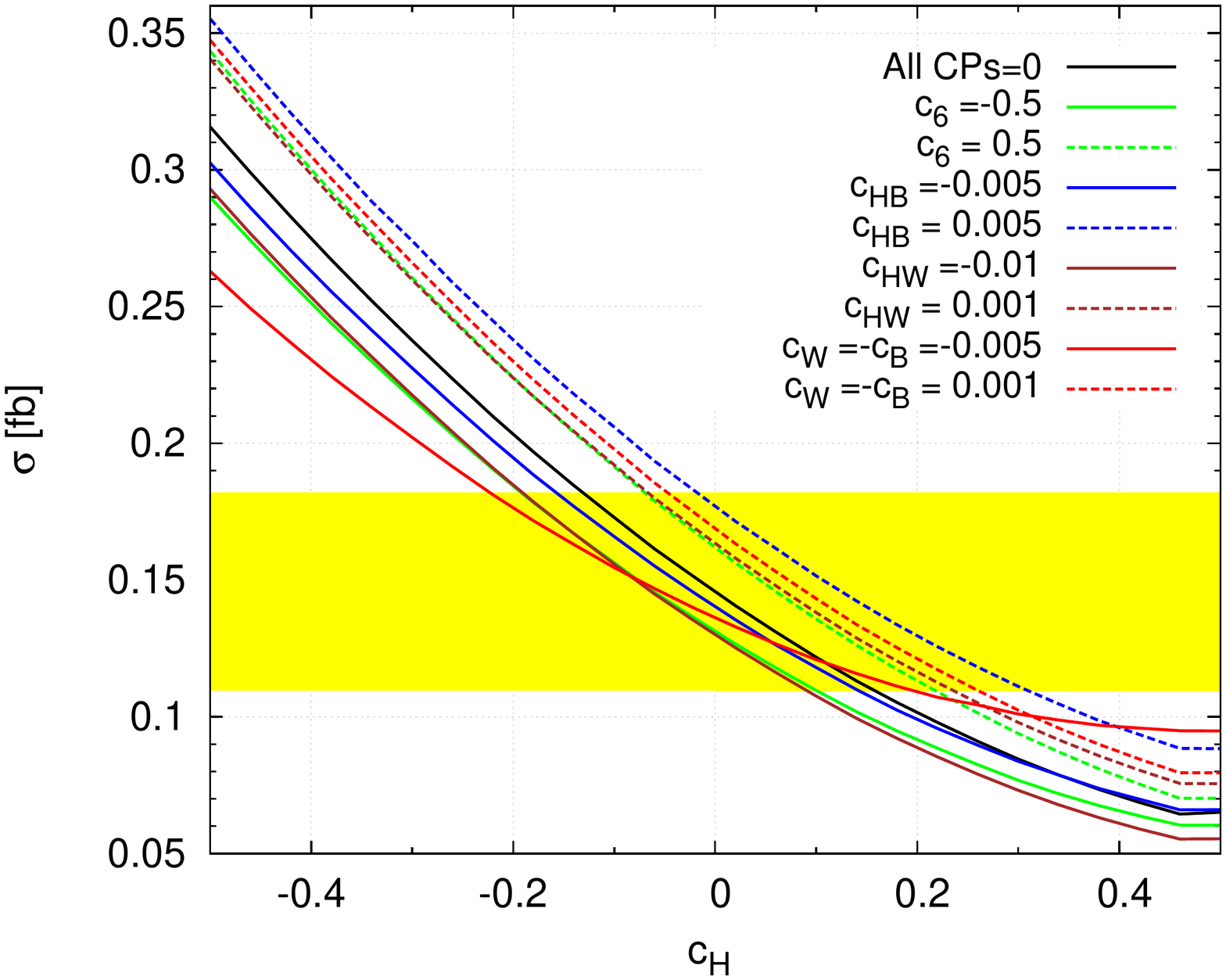}  \\
\end{tabular}
\caption{
Cross section of $ZHH$ production against $c_6$ (left) and $c_H$ (right), when some of the other selected relevant parameters assume typical values is compared against the case when only $c_6$ or $c_H$  is present. The black solid lines corresponds to the case when all parameters other than $c_6$ (left) or $c_H$ (right) vanish. The centre of mass energy is assumed to be 
$\sqrt{s}=800$ GeV. In each case, all other parameters are set to zero. The yellow band indicates the $3\sigma$ limit 
of the SM cross section, with integrated luminosity of 1000 fb$^{-1}$.}
\label{fig:cs_c6cH}
\end{figure}

We consider the influence of $c_6$ on the cross section in Fig.~\ref{fig:cs_c6cH} (left). We have compared the variation of cross section  with $c_6$ keeping all other parameters to the SM value, with the cases when some of the relevant parameters having non-standard values. The $3\sigma$ region (yellow band) of the SM value of the cross section, considering an integrated luminosity of 1000 fb$^{-1}$, is presented in these plots so as to make an estimate of the reach on the $c_6$. The plots clearly indicate the correlation between the influence of different parameters on the cross section. For example, assuming only $c_6$ takes a non-zero value, the reach at $3\sigma$ level is approximately $-0.5<c_6<0.4$, as indicated by the black solid line. However, as indicated by the red solid line, if we assume a typical value of $c_W=-c_B=-0.005$, the lower limit is considerably relaxed, with some moderate change in the upper bound to 0.5. On the other hand, for the case with $c_W=-c_B=0.001$, where the sign is reveresed, the upper bound becomes more stringent, whereas the lower bound is more relaxed. A similar story can be read out for the cases with the presence of other parameters as well. The effect of all the parameters $c_W,~~ c_{HW}$ and $c_{HB}$, which contribute to the $ZZH$ and $ZZHH$ couplings are found to be significant. Strong dependence of the sensitivity of $c_6$ on the presence of $c_H$ is somewhat expected, for both parameters contribute to the $HHH$ coupling. In Fig.~\ref{fig:cs_c6cH} (right), similarly, we consider the variation of the cross section with $c_H$, again exploring the effect of different parameters on it. Here again, the depndence on all the parameters on the sentivity of $c_H$ on the cross section is found to be significant for chosen typical values of the parameters.

\begin{figure}[h]\centering
\begin{tabular}{c}
\hspace{-8mm}
\includegraphics[angle=0,width=82mm]{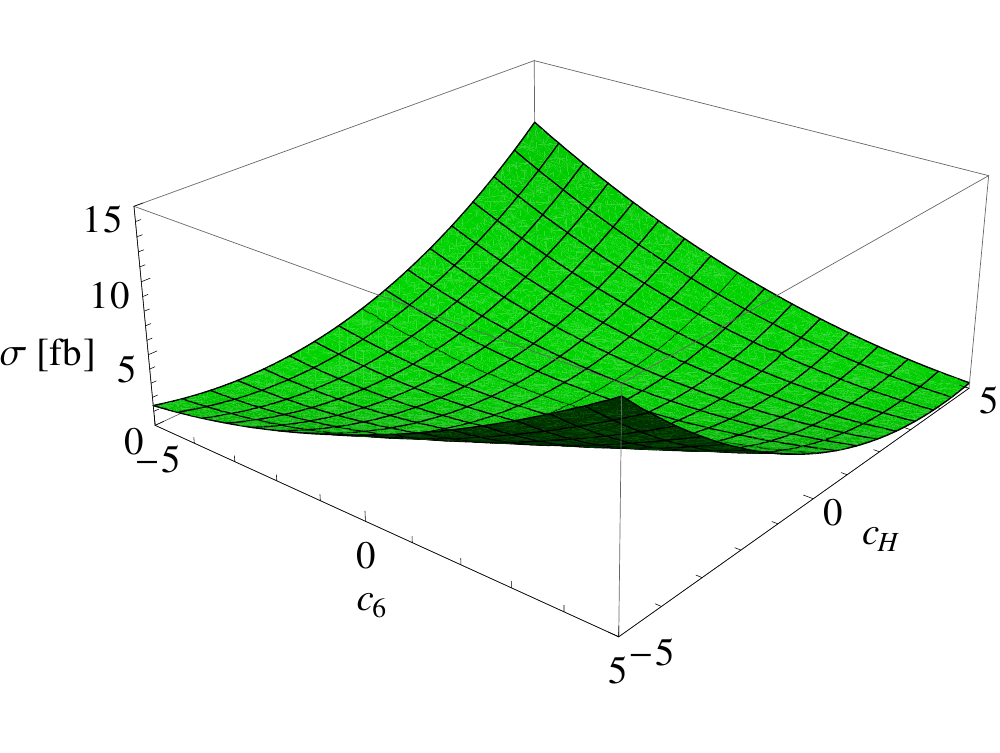}
\end{tabular}
\caption{Cross section of $ZHH$ production plotted against $c_6$ and $c_H$ at $\sqrt{s}=800$ GeV, with all other parameters set to zero.}
\label{fig:sig_2param_c6cH}
\end{figure}

In Fig.~\ref{fig:sig_2param_c6cH}, the cross section is plotted against $c_6$ and $c_H$. The correlation of the sensitivity between the two parameters is clear. The opposite sign combination seems to be more sensitive to the cross section, and therefore more stringent constraints could be drawn in this case compared to the same sign case.

\begin{figure}[h]\centering
\begin{tabular}{c c}
\hspace{-8mm}
\includegraphics[angle=0,width=80mm]{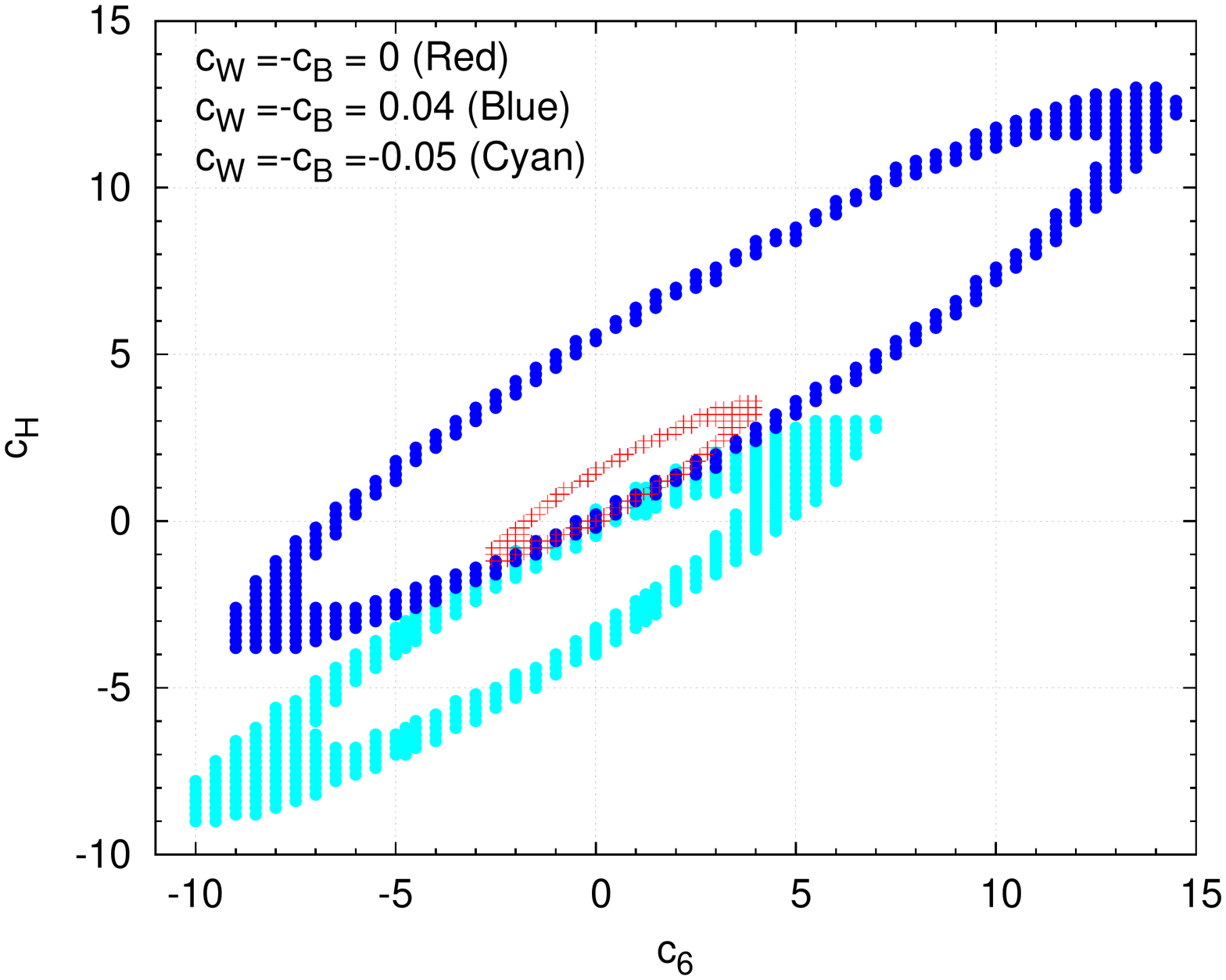} &
\includegraphics[angle=0,width=80mm]{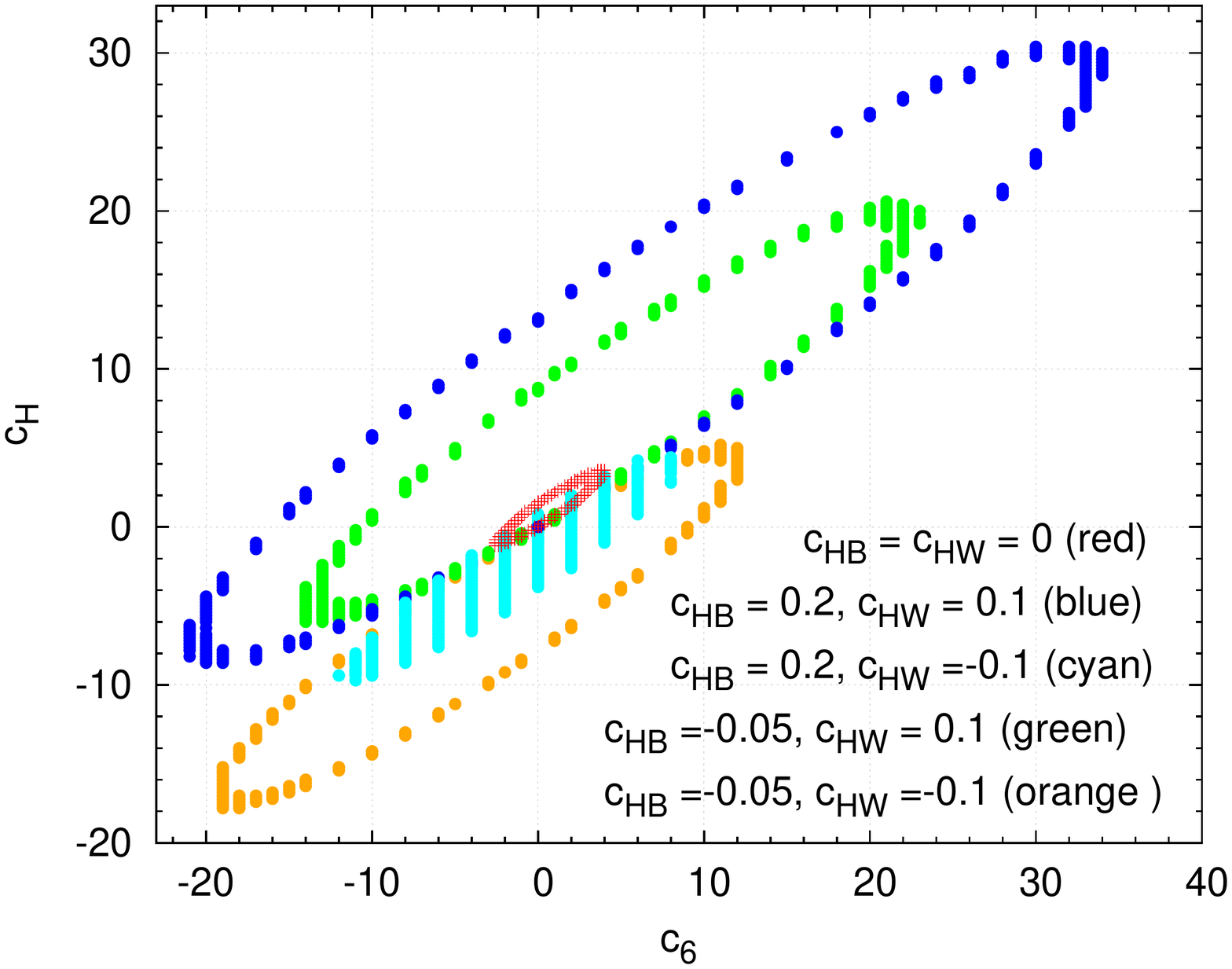}\\
\end{tabular}
\caption{
The shaded regions correspond to regions in the $c_6$-$c_H$ plane with the total cross section is within the $3\sigma$ limit when $c_6=c_H=0$ in each case, for an integrated luminosity of 1000 fb$^{-1}$ at a centre of mass energy of 800 GeV. Values of the other anomalous couplings are as indicated in the figure, with all other couplings set to zero.}
\label{fig:lt_c6_cH}
\end{figure}

The reach of ILC on the trilinear Higgs coupling through the process being considered can be established by considering the $3\sigma$ limit of the cross section at an integrated luminosity of 1000 $fb^{-1}$ as presented in Fig.~\ref{fig:lt_c6_cH}, for the case of SM, and cases with non-vanishing anomalous $ZZH$ and $ZZHH$ couplings. Please note that, when cross section is considered as a function of $c_6$ and  $c_H$,  the result is a second order polynomial with these two parameters. With this, the $3\sigma$ limit of the cross section leads to an elliptic equation correponding to the relation between these two parameters. This result in an elliptic band in the $c_6 - c_H$ plane respecting the $3\sigma$ limit of the cross section. As is evident from the plots, these allowed bands of the parameters move in the parameter space, depending on the values of the other parameters, as illustrated by the cases of $c_W=-c_B$, $c_{HW}$ and $c_{HB}$. These results also illustrate how important the signs of different couplings are in a study of the sensitivity of the trilinear Higgs couplings. What we may learn from the above is  that the limits drawn with assuming the absence of all other parameters may not depict the actual situation. 

It is important to know the behaviour of the kinematic distributions, and how the anomalous parameters influence these, to derive any useful and reliable conclusions from the experimental results. This is so, even in cases where the fitting to obtain the reach of the parameters is done with the total number of events, as the reconstruction of events and the reduction of the background depend crucially on the kinematic distributions of the decay products. In the following we shall present some illustrative cases of distributions at the production level, in order to understand the effect of different couplings on these. The changes in the kinematic distributions at the production level will also be carried over to the distributions of their decay products. Presently we woudl like to be content with the analysis at the production level, considering the limited scope of this work. As mentioned earlier we shall focus on an ILC running at a centre of mass energy of 800 GeV for our study.

\begin{figure}[h]\centering
\begin{tabular}{c c }
\hspace{-10mm}
\includegraphics[angle=0,width=80mm]{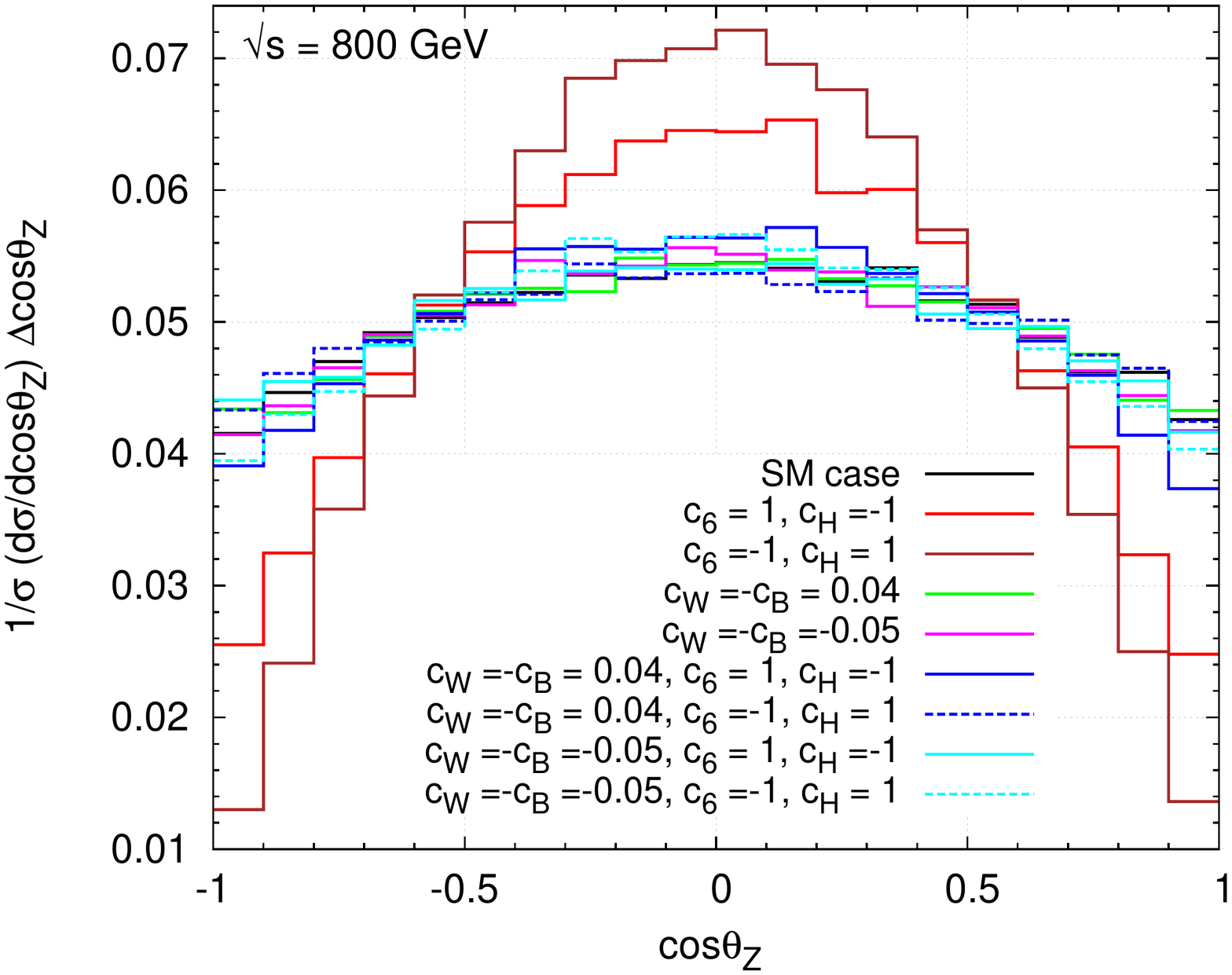} &
\includegraphics[angle=0,width=80mm]{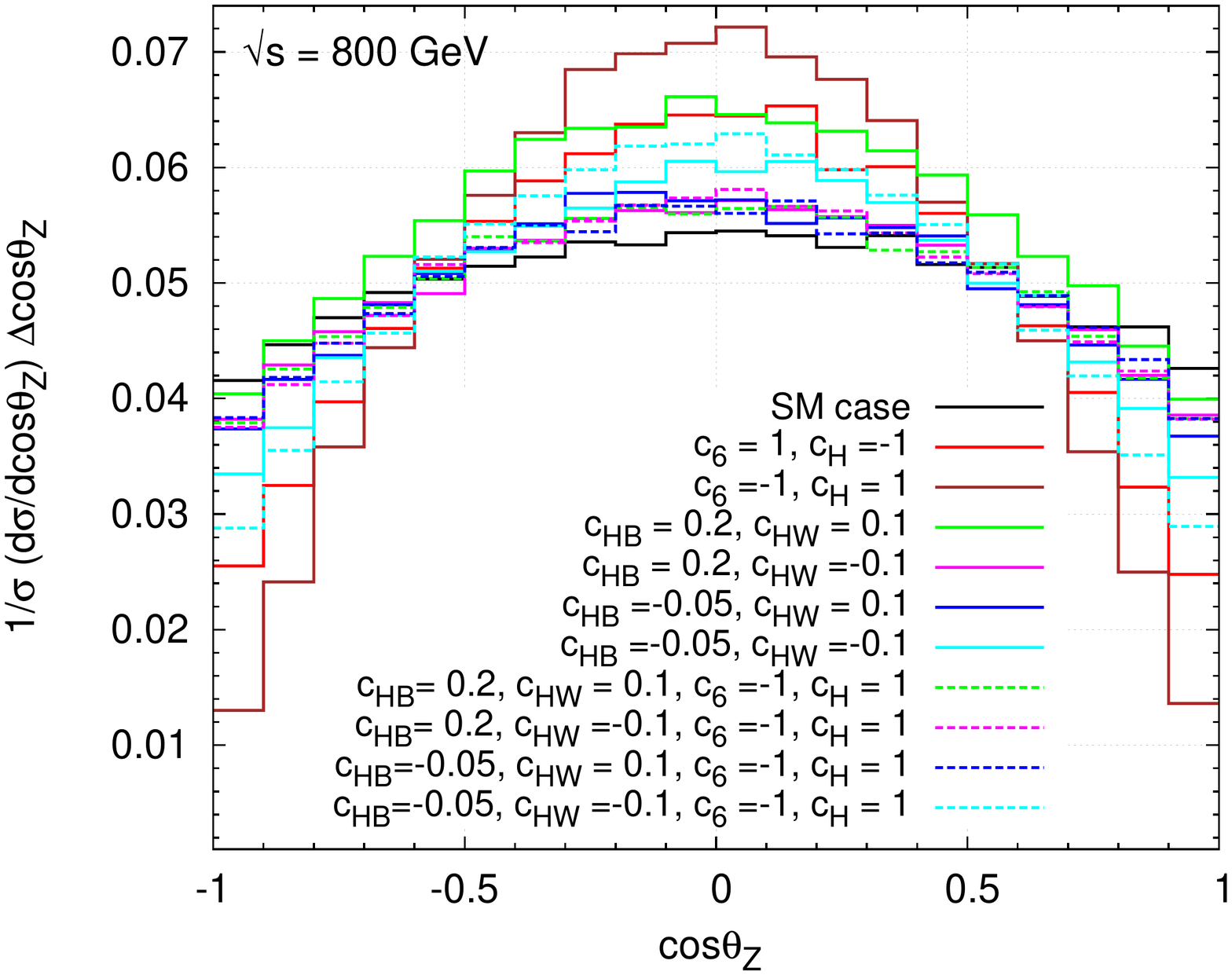}\\
\hspace{-10mm}
\includegraphics[angle=0,width=80mm]{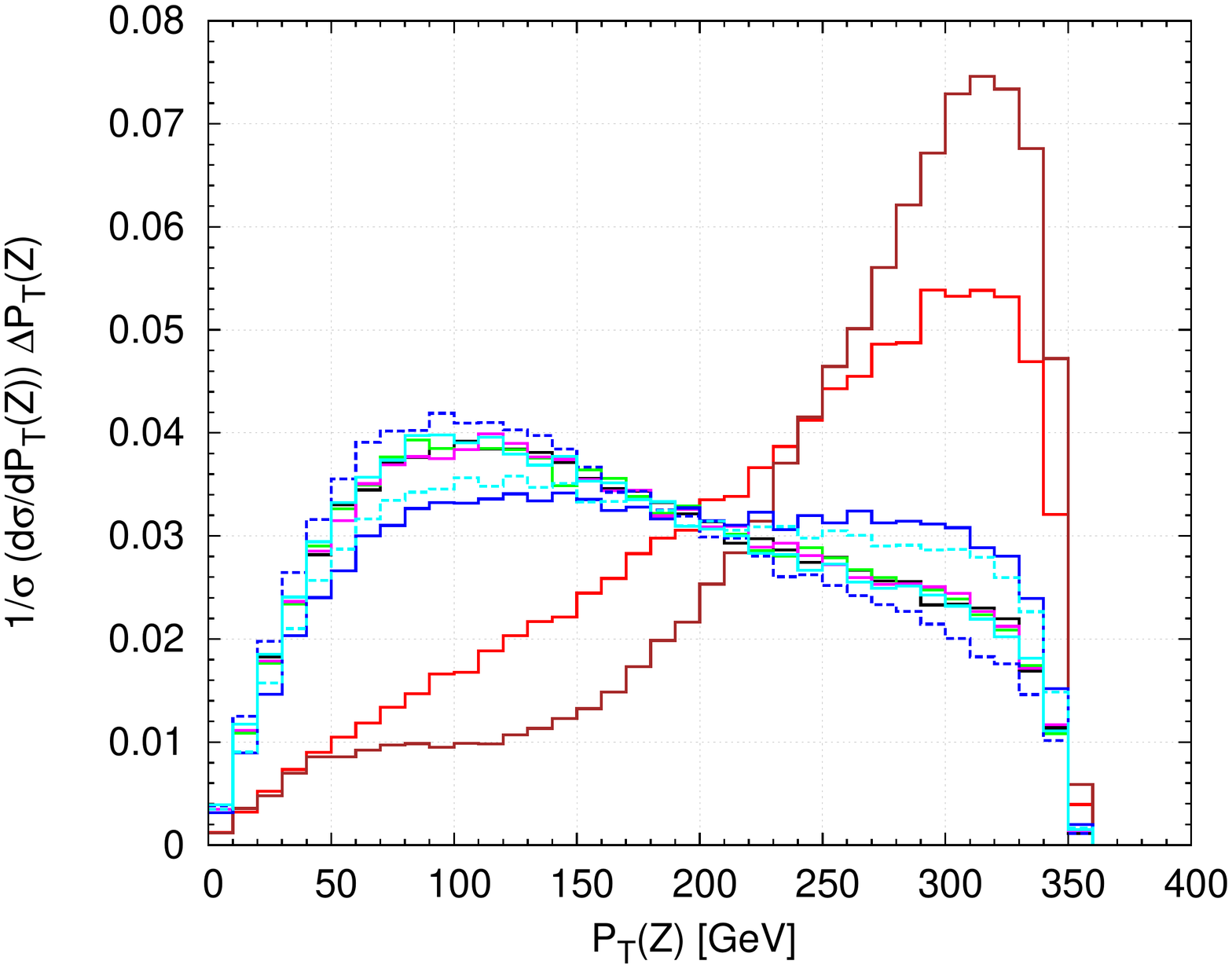} &
\includegraphics[angle=0,width=80mm]{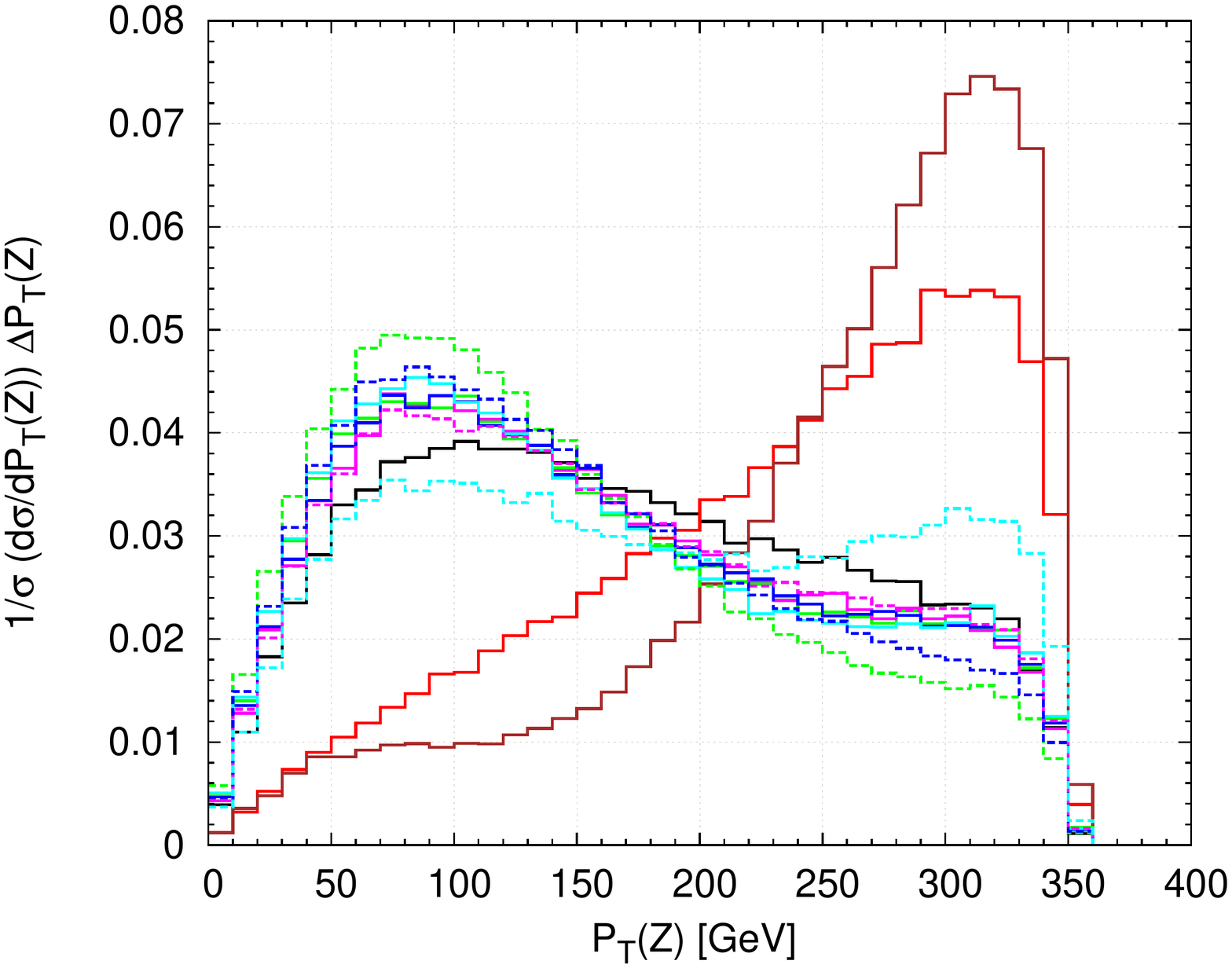}\\
\hspace{-10mm}
\includegraphics[angle=0,width=80mm]{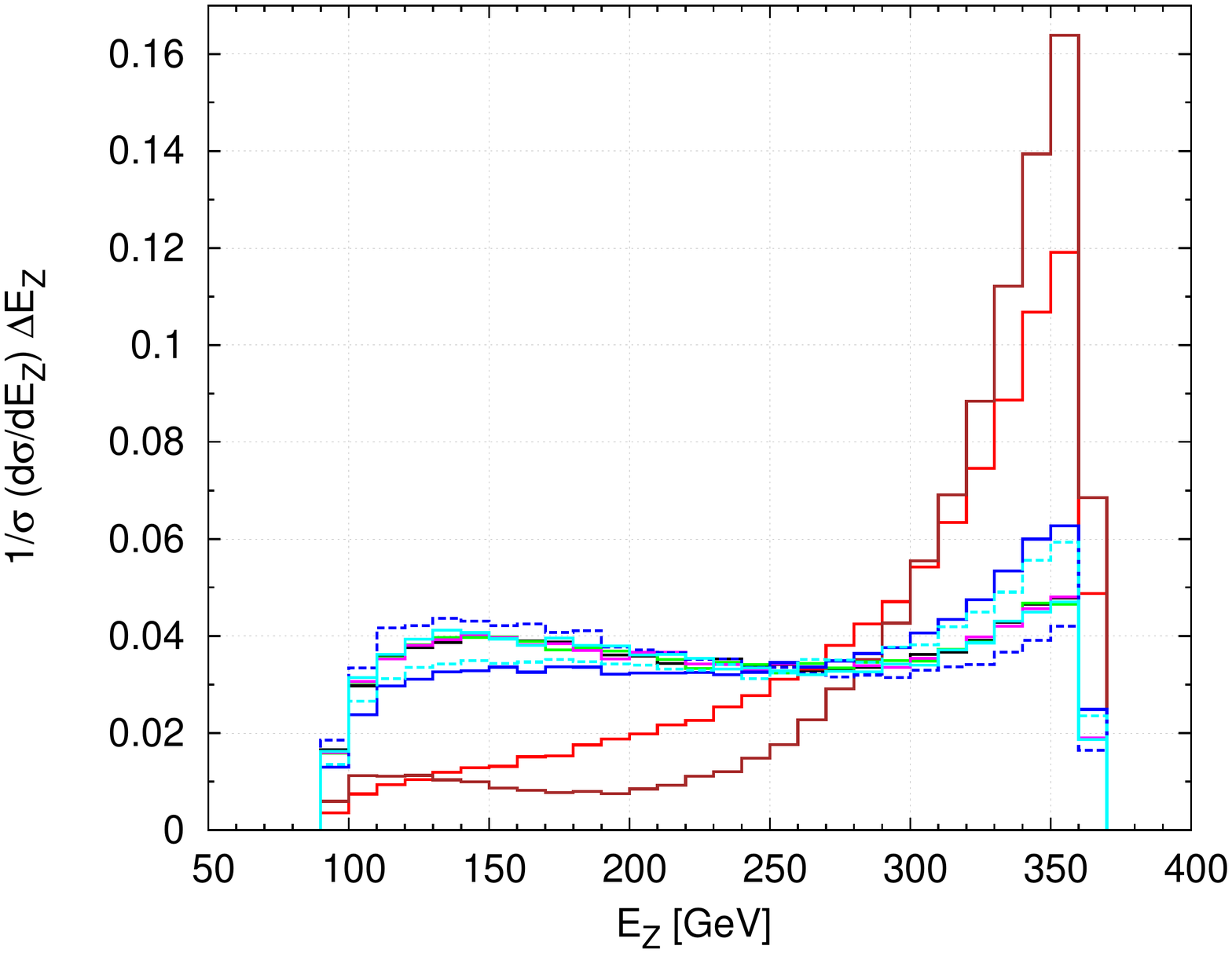} &
\includegraphics[angle=0,width=80mm]{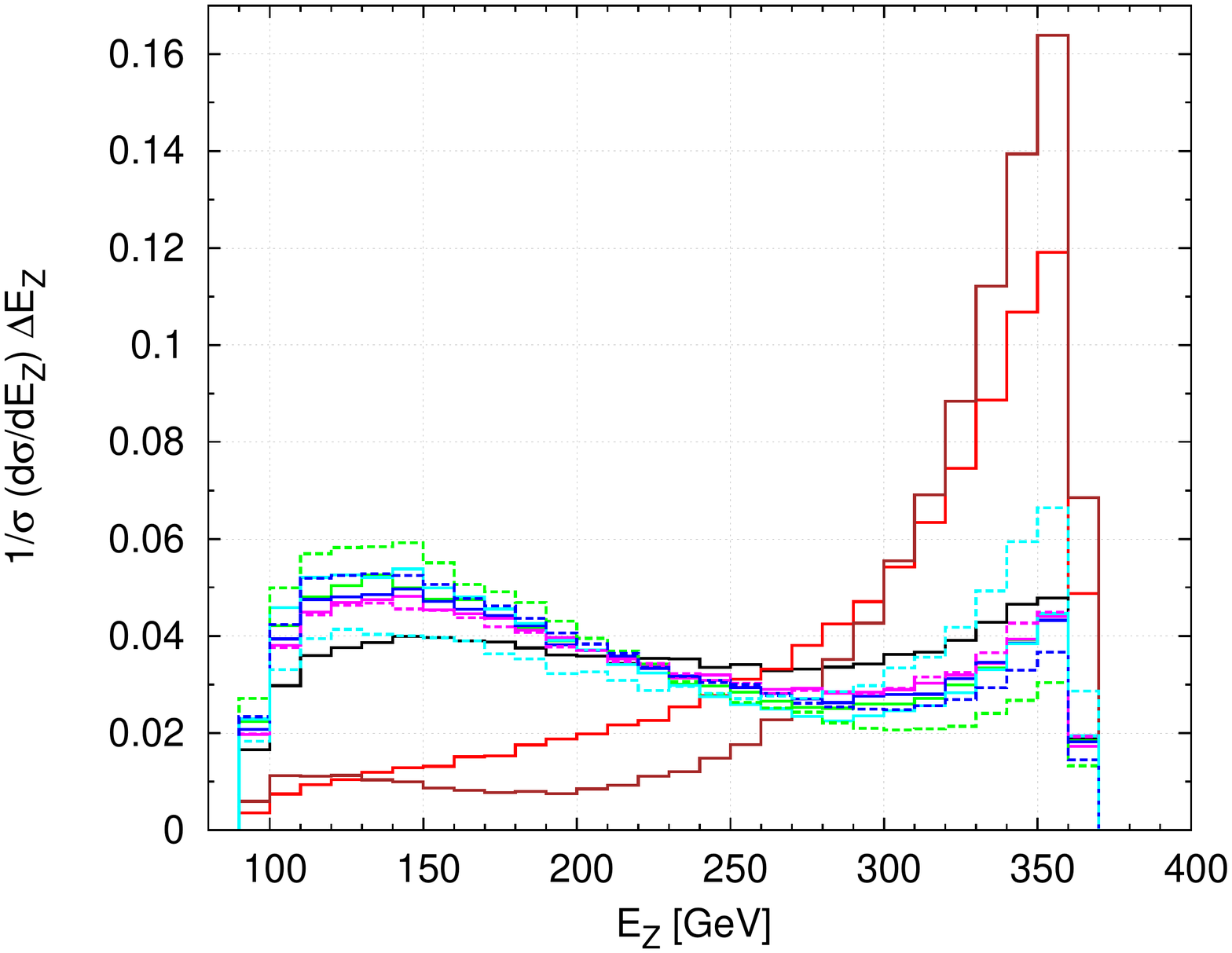} \\
\end{tabular}
\caption{Distributions of the 
$\cos\theta_Z$, Transverse Momentum, and Energy of the $Z$ boson for the anmalous coupling values as in the inset, illustrating how the presence of $c_W$ (first column), and  $c_{HW}$ and $c_{HB}$ (second column) affect the influence of $c_6$ and $c_H$. A centre of mass energy of  $800$ GeV is assumed.}
\label{fig:ctZ_ptZ_EZ}
\end{figure}

We first consider in Fig.\ref{fig:ctZ_ptZ_EZ} (top row), the normalized $\cos\theta_Z$ distributions of the $Z$ boson for the case of SM, as well as for different cases with anomalous couplings. The normalized distributions presents the difference in the shape, which brings out the qualitative difference in a more visible manner. The figure on the left shows the case with $c_{W}=-c_B$  taking typical values, while the other parameters set to zero, whereas the figure on the right considers $c_{HW}$ and $c_{HB}$ non-zero, while setting other parameters to zero . The case with only $c_6$ and $c_H$ taking non-zero values, when compared with the SM case shows a perceivable  change in the distribution with more number of events piling in the small $\cos\theta_Z$ region. Such an experimental observations could therefore be considered as an indication of the anomalous $HHH$ coupling. On the other hand, the presence of anomlaous $c_W$ and $c_B$ couplings does not affect the distribution much. More importantly, in their presence, the non-zero $c_W$ and $c_B$, the distribution remains close to the SM distribuiton, even with non-zero $c_6$ and $c_H$. Thus, a conclusion regarding the presence or otherwise of the $HHH$ coupling drawn from the $\cos\theta_Z$ distribution will depend on the values of $c_W$ and $c_B$. The figure on the right tells a similar story for the case of $c_{HW}$ and $c_{HB}$ replaceing $c_W$.  In Fig.\ref{fig:ctZ_ptZ_EZ} (second row) and (third row), the $p_T$ and energy distributions of the $Z$ boson are plotted. Here too, we see that if only $c_6$ and $c_H$ are considered to be non-zero, events with high $p_T$ and high energy $Z$ bosons are preferred much more in comparison with the SM case. This conclusion is upset with the simultaneous presence of other parameters related to  $ZZH$ coupling. 
\begin{figure}[h]\centering
\begin{tabular}{c c}
\hspace{-10mm}
\includegraphics[angle=0,width=80mm]{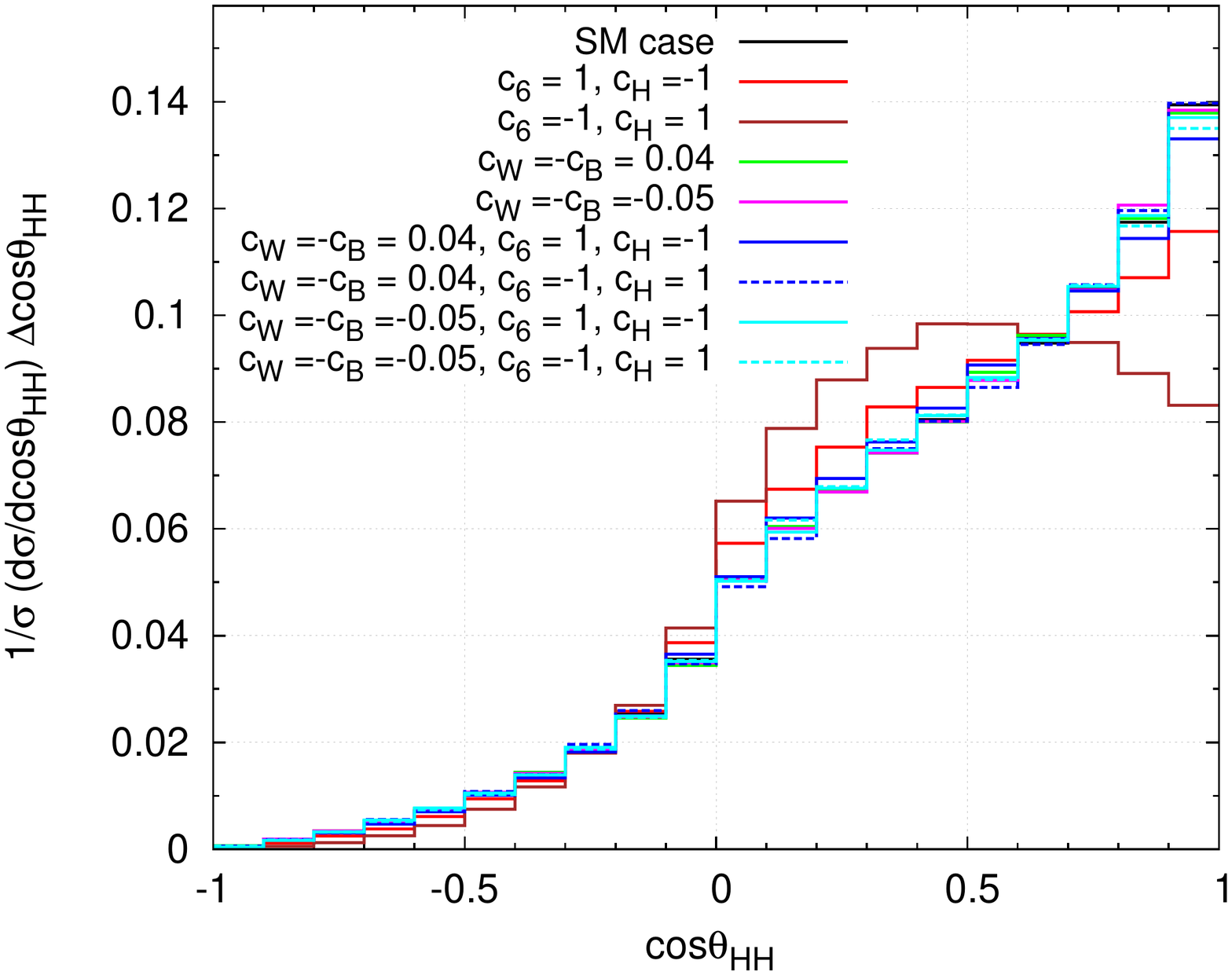}&
\includegraphics[angle=0,width=80mm]{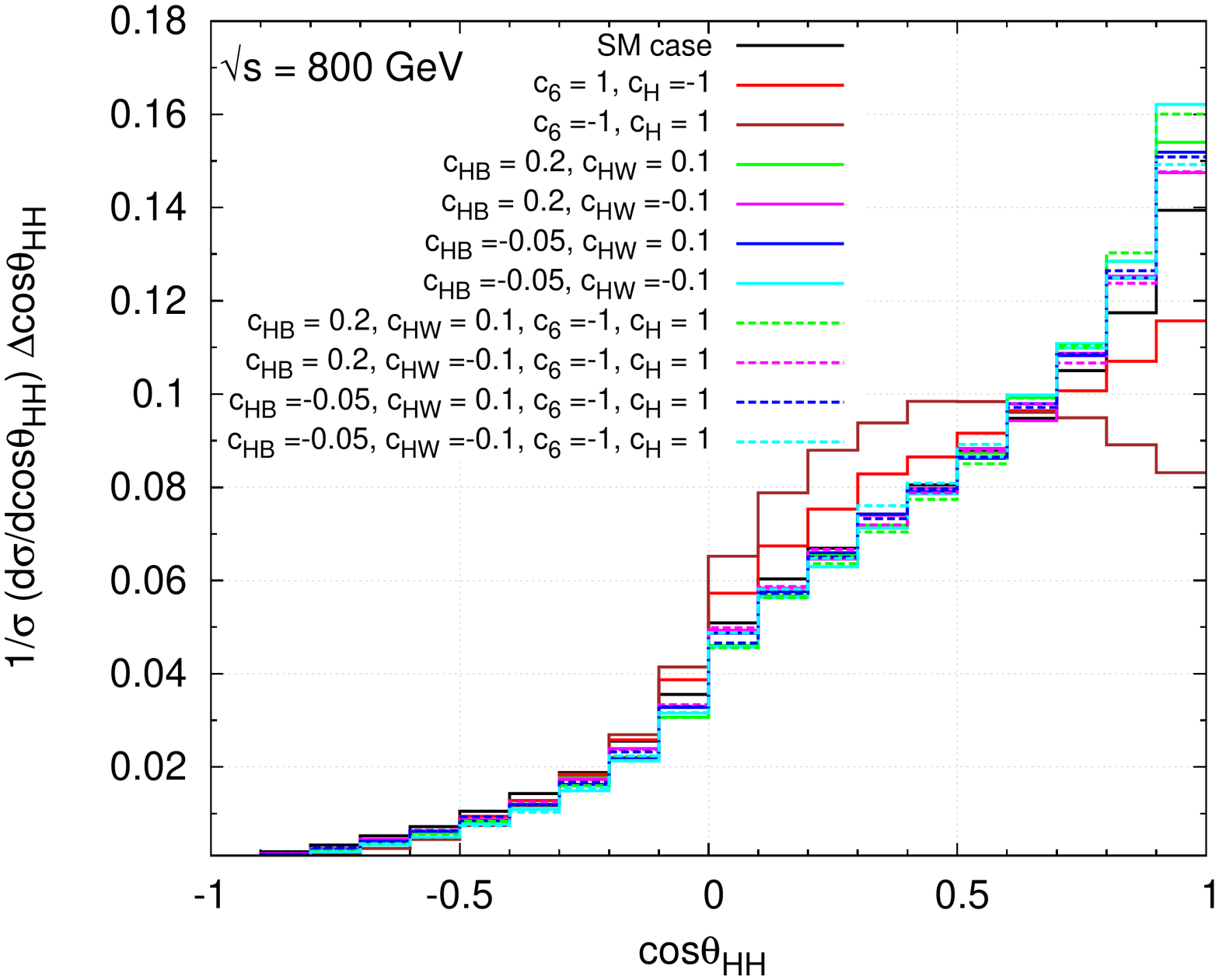}  \\
\hspace{-10mm}
\includegraphics[angle=0,width=80mm]{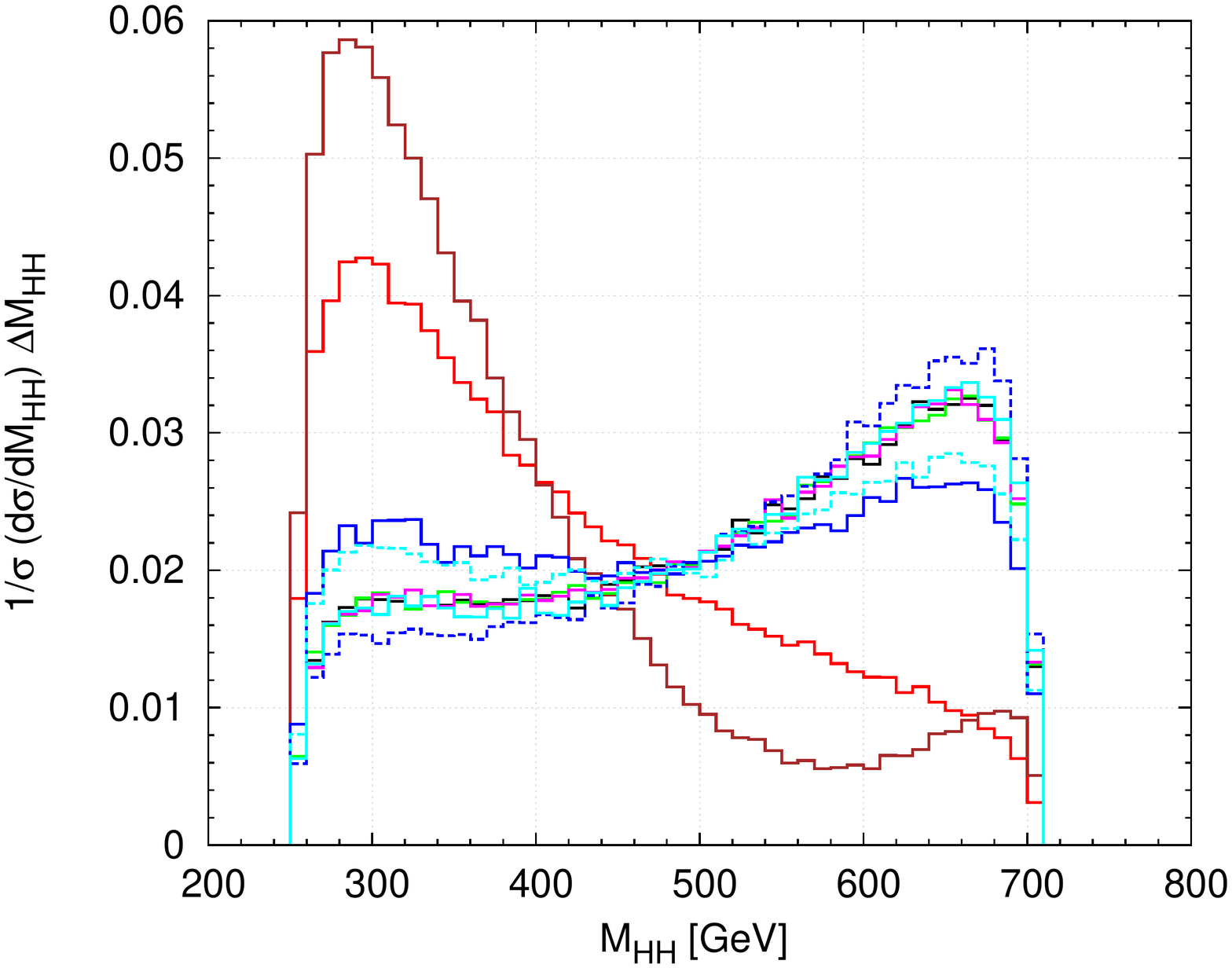} &
\includegraphics[angle=0,width=80mm]{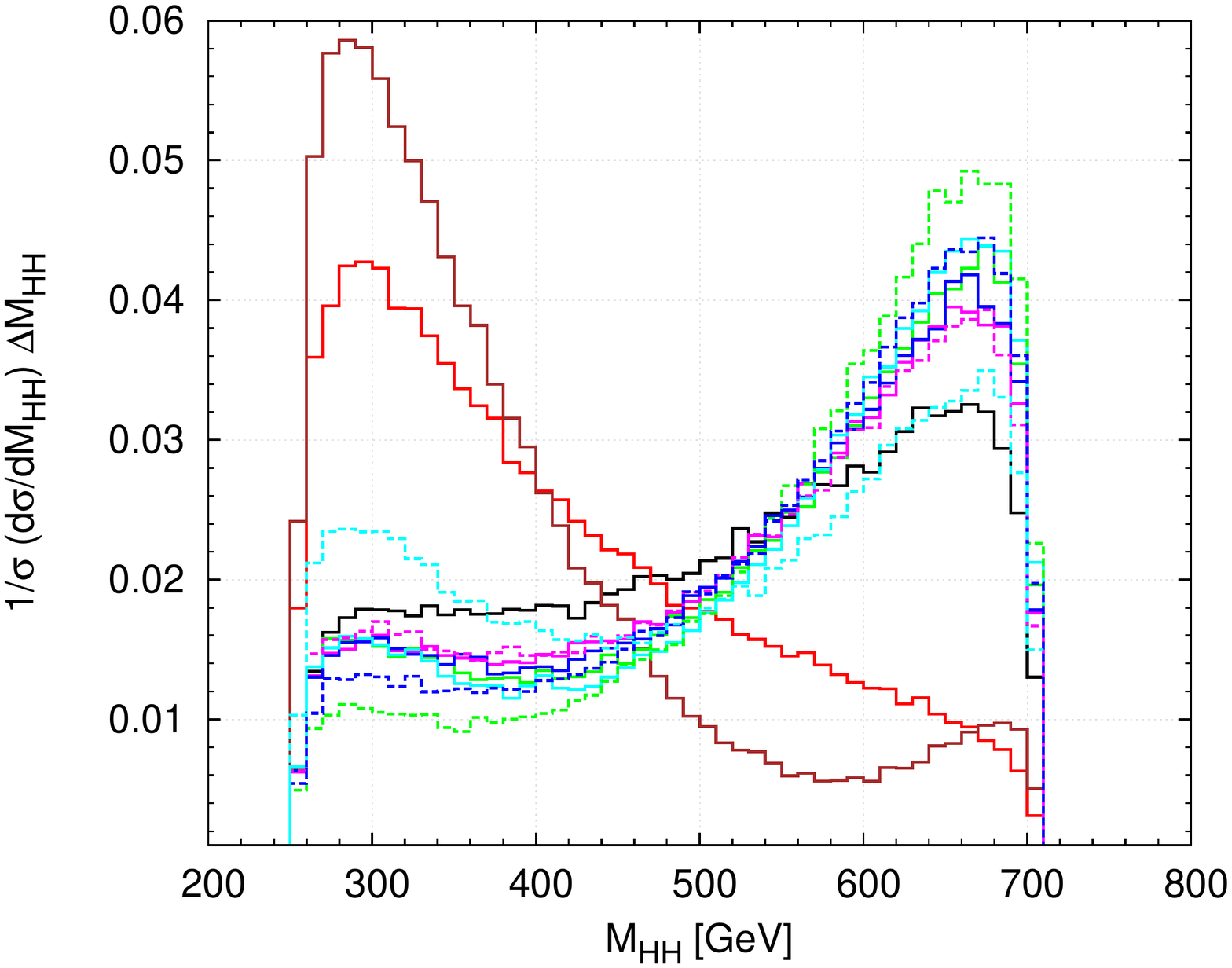}
\end{tabular}
\caption{$\cos\theta_{HH}$ and the invariant mass of $HH$ distributions for the anmalous coupling values as in the inset, illustrating how the presence of $c_W$ (first column), and  $c_{HW}$ and $c_{HB}$ (second column) affect the influence of $c_6$ and $c_H$. A centre of mass energy of  $800$ GeV is assumed.}
\label{fig:ct_m_HH}
\end{figure}
The distribution of the opening angle between the two Higgs bosons as well as their invariant mass distribution presented in Fig.~\ref{fig:ct_m_HH} indicate the same feature captured in the various distributions of the $Z$ bosons. While in all cases including the SM case, most of the events are in the forward hemisphere, in the presence of non-vanishing $c_6$ and $c_H$, but with $c_W=c_{HW}=c_{HB}=0$, the events are more evenly distributted within the forward hemispehere, compared to the rest of the cases including the SM case. The $HH$ invariant  mass demonstrate an even more dramatic difference in the different cases mentioned above. 

The conclusions that we draw from the above considerations is that single parameter considerations to understand the effect of $HHH$ coupling will not be realistic, if other relevant gauge-Higgs couplings receive anomalous contributions. Our preliminary investigation clearly indicates that the correlations can be rather strong, for all the relevant parameters, and one need to consider  a careful analysis to obtain realistic limits on the parameters.

\subsection{$e^+e^-\rightarrow HH\nu\bar \nu$ process}

\begin{figure}[!t]\centering
\begin{tabular}{c c}
\includegraphics[angle=0,width=100mm]{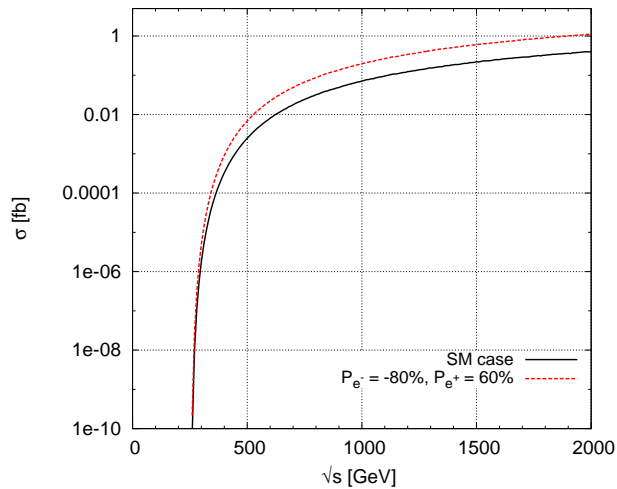} 
\end{tabular}
\caption{Total cross section of  $e^- e^+ \rightarrow \nu_e \bar\nu_e H H$ in the case of unpolarized and polarized beams, as indicated.}
\label{fig:sig_rs_nnhh}
\end{figure}

We shall now turn our attention to the second process involving $HHH$ couplings, as well as gauge-Higgs couplings. We consider the two Higgs production with missing energy through the process $e^+ e^- \rightarrow HH\nu \bar \nu$. The previous process, $e^+ e^- \rightarrow HHZ$, with $Z\rightarrow \nu \bar \nu$ has the same final state. But, this can be easily separated from the rest of the contributions due, in the SM, to the channels presented in the Feynman diagrams given in Fig.~\ref{fig:fdnnhh}, through, for example considering the missing invariant mass. The cross section for the process is plotted against the centre of mass energy for the case of polarized as well as unpolarized beams in Fig.~\ref{fig:sig_rs_nnhh}. The advantage of very high energy collider is evident here. We shall consider a centre of mass energy of 2 TeV, for which the cross section is close to 0.4 fb in case of unpolarized beams, and slightly more than 1 fb for $e^-$ beam of $-80\%$ polarization and $e^+$ beam with $+60\%$ polarization. This study will complement the the study of the $ZHH$ production in the sense that the physical couplings involved are $HHH$ along with $WWH$ and $WWHH$ instead of the ones involving the neutral gauge bosons. Although in the language of the effective Lagrangian, the couplings involved are similar to the ones in the previous process, their involvement in the current process is expected to be different.

\begin{figure}[h]\centering
\begin{tabular}{c c}
\hspace{-5mm}
\includegraphics[angle=0,width=80mm]{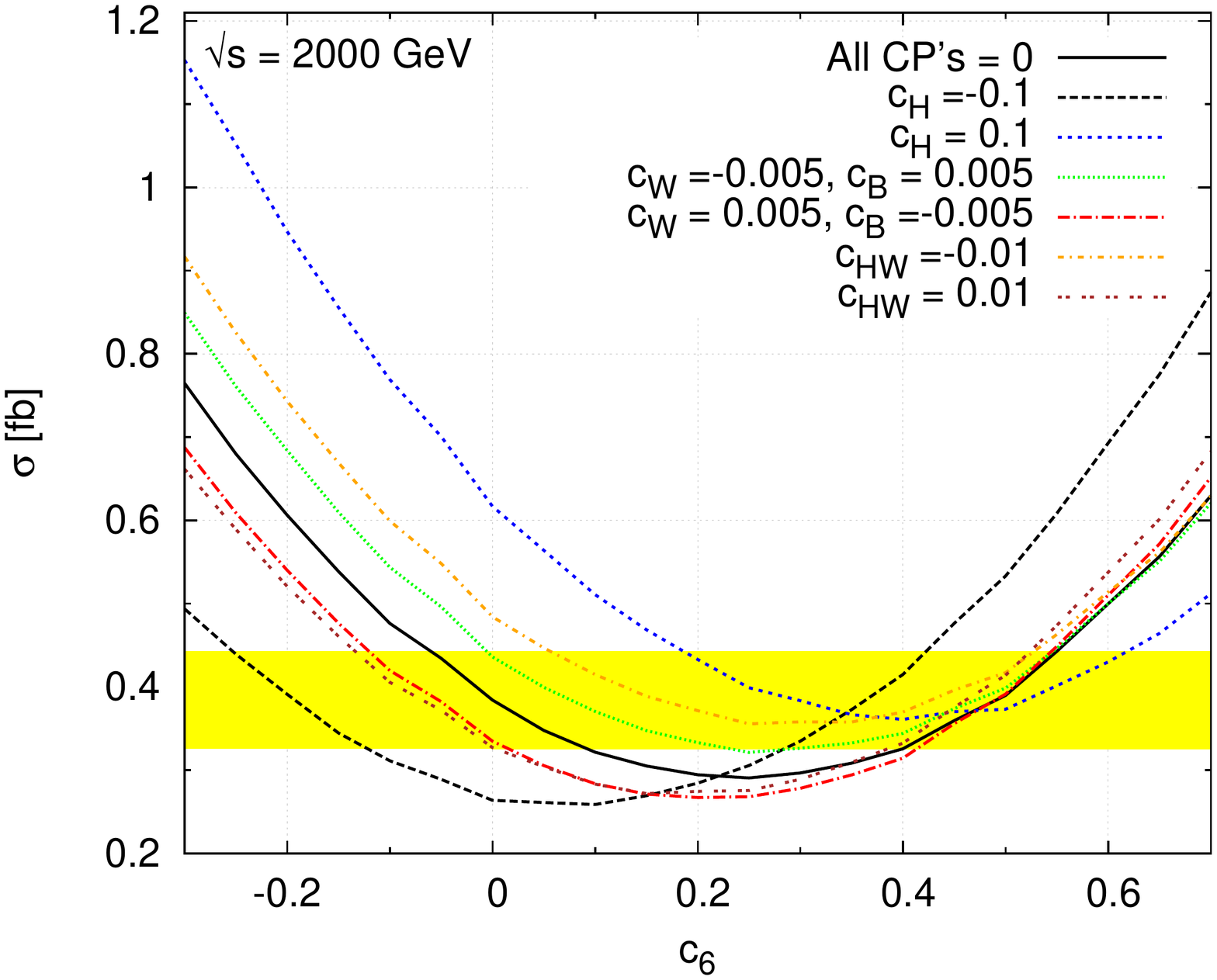}&
\includegraphics[angle=0,width=80mm]{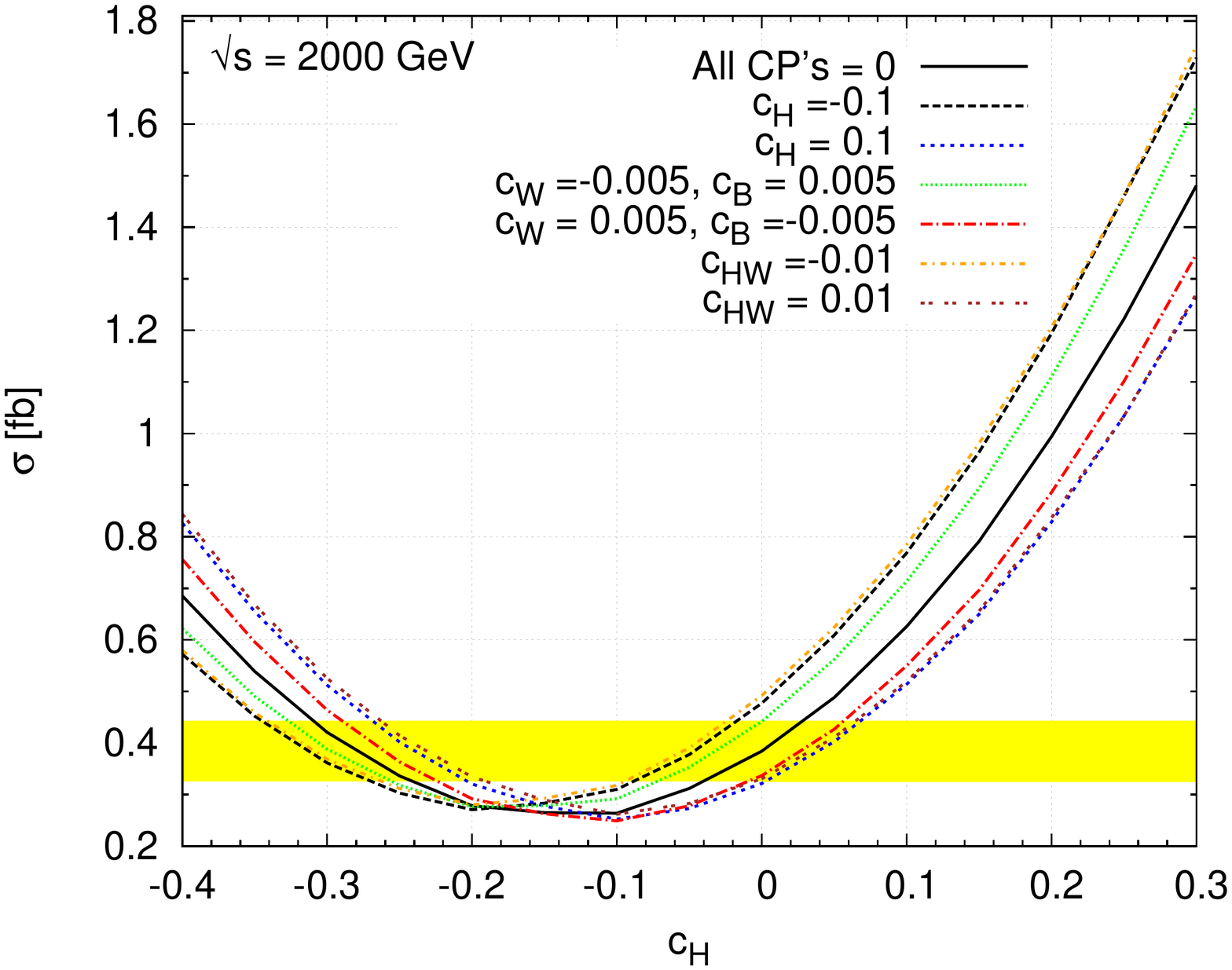}  \\
\end{tabular}
\caption{
Cross section of $\nu\bar\nu HH$ production against $c_6$ (left) and $c_H$ (right), when some of the other selected relevant parameters assume typical values is compared against the case when only $c_6$ or $c_H$  is present. The black solid lines corresponds to the case when all parameters other than $c_6$ (left) or $c_H$ (right) vanish. The centre of mass energy is assumed to be 
$\sqrt{s}=2$ TeV. In each case, all other parameters are set to zero. The yellow band indicates the $3\sigma$ limit 
of the SM cross section.}
\label{fig:cs_c6_nnhh}
\end{figure}

As in the earlier case, the sensitivity of $c_6$ and $c_H$ on the total cross section at the centre of mass energy of 2 TeV is presented in Figs.~\ref{fig:cs_c6_nnhh}, when all other parameters are set to zero, as well as in the presence of some of the relevant parameters. We have included the $3\sigma$ band of the SM cross section assuming 1000 fb$^{-1}$ luminosity. Clearly, the correlation is perceivable, and the conclusions are similar to the case of  $ZHH$ production, that the sensitivity of $HHH$ coupling on the process considered strongly depend on the values of other parameters relevant to $WWH$ and $WWHH$ couplings. 

\begin{figure}[!t]\centering
\begin{tabular}{c c}
\hspace{-8mm}
\vspace{-5mm}
\includegraphics[angle=0,width=80mm]{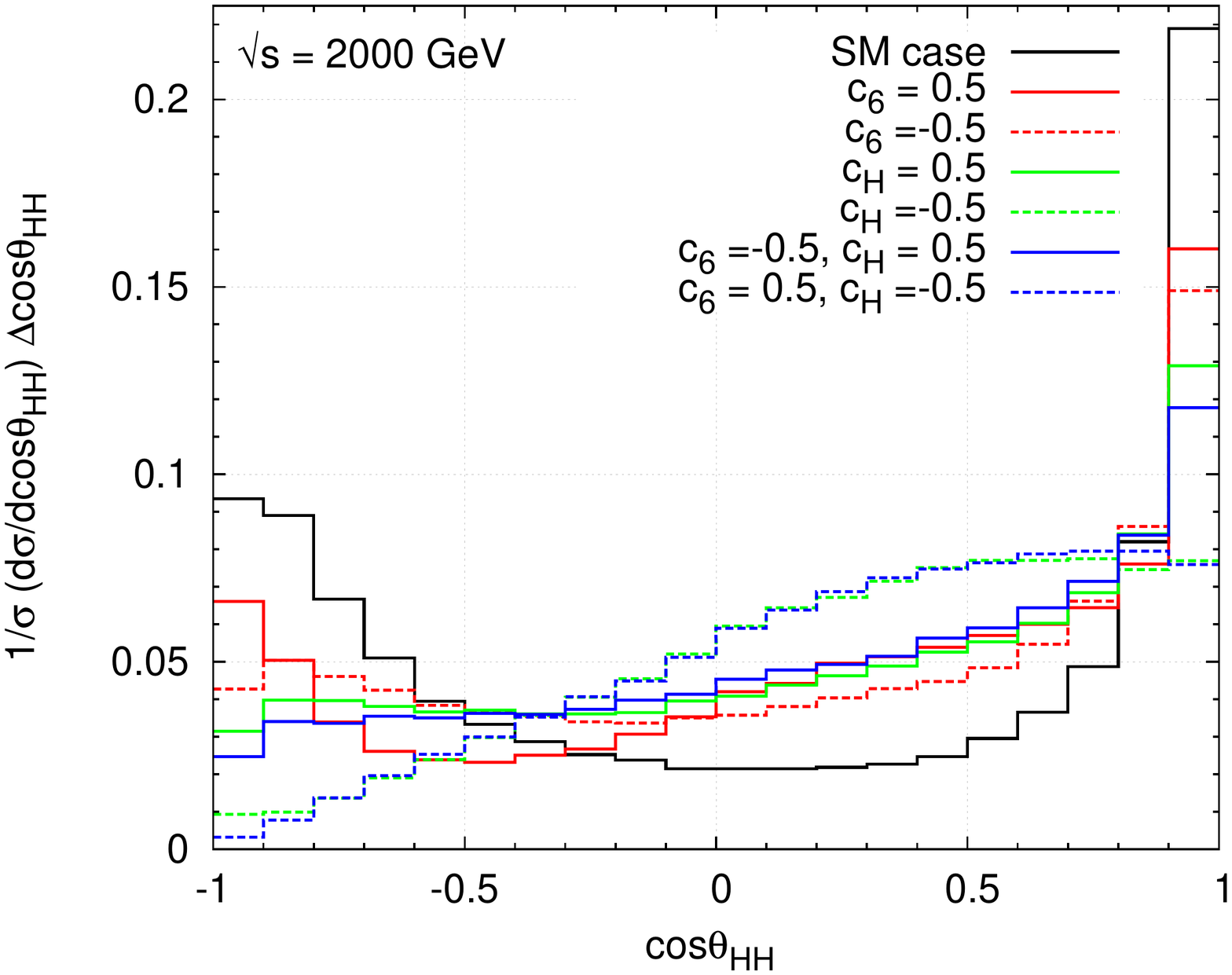} &
\hspace{-5mm}
\vspace{-5mm}
\includegraphics[angle=0,width=80mm]{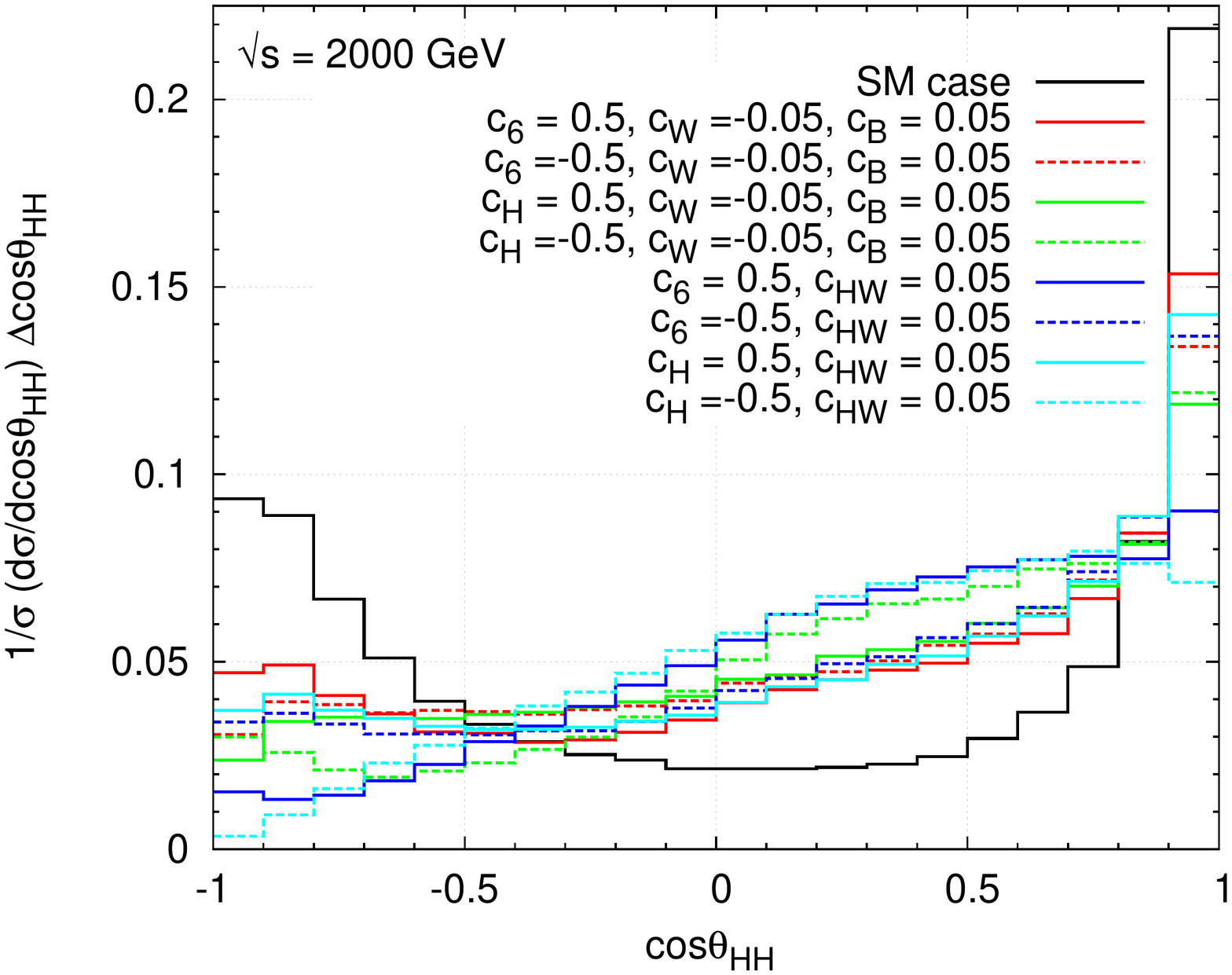} \\
\hspace{-8mm}
\vspace{-5mm}
\includegraphics[angle=0,width=80mm]{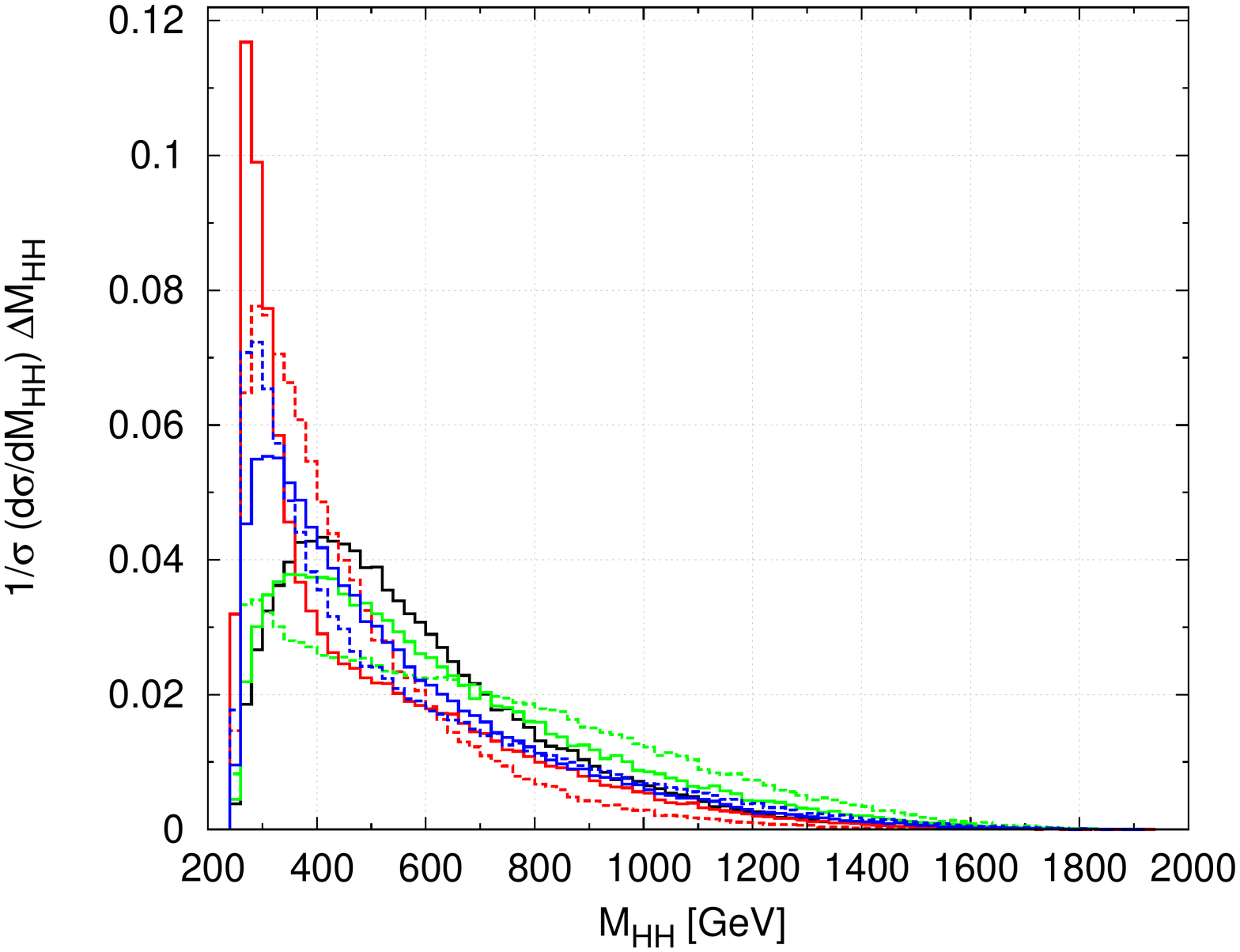} &
\hspace{-5mm}
\vspace{-5mm}
\includegraphics[angle=0,width=80mm]{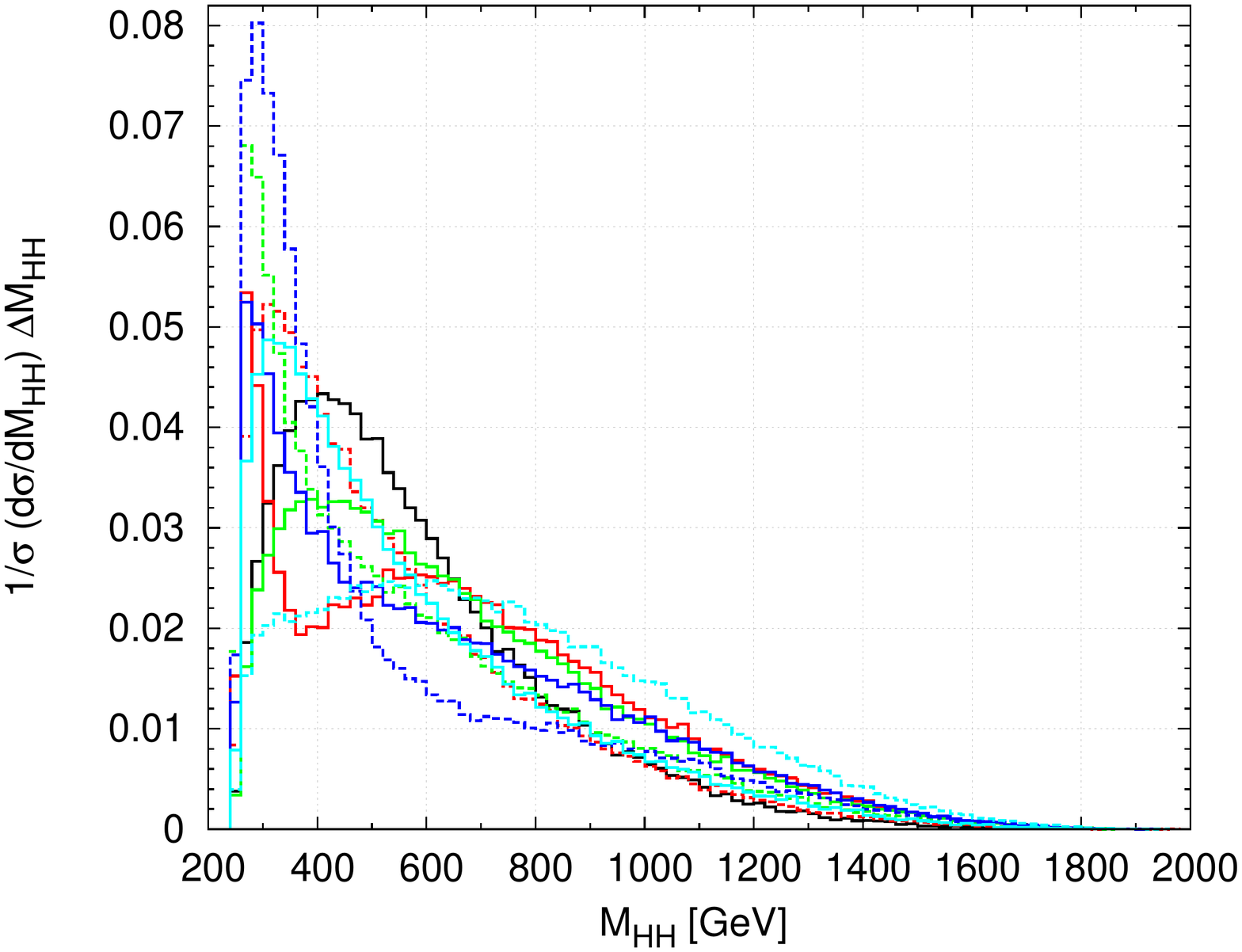}\\
\hspace{-5mm}
\vspace{-5mm}
\includegraphics[angle=0,width=80mm]{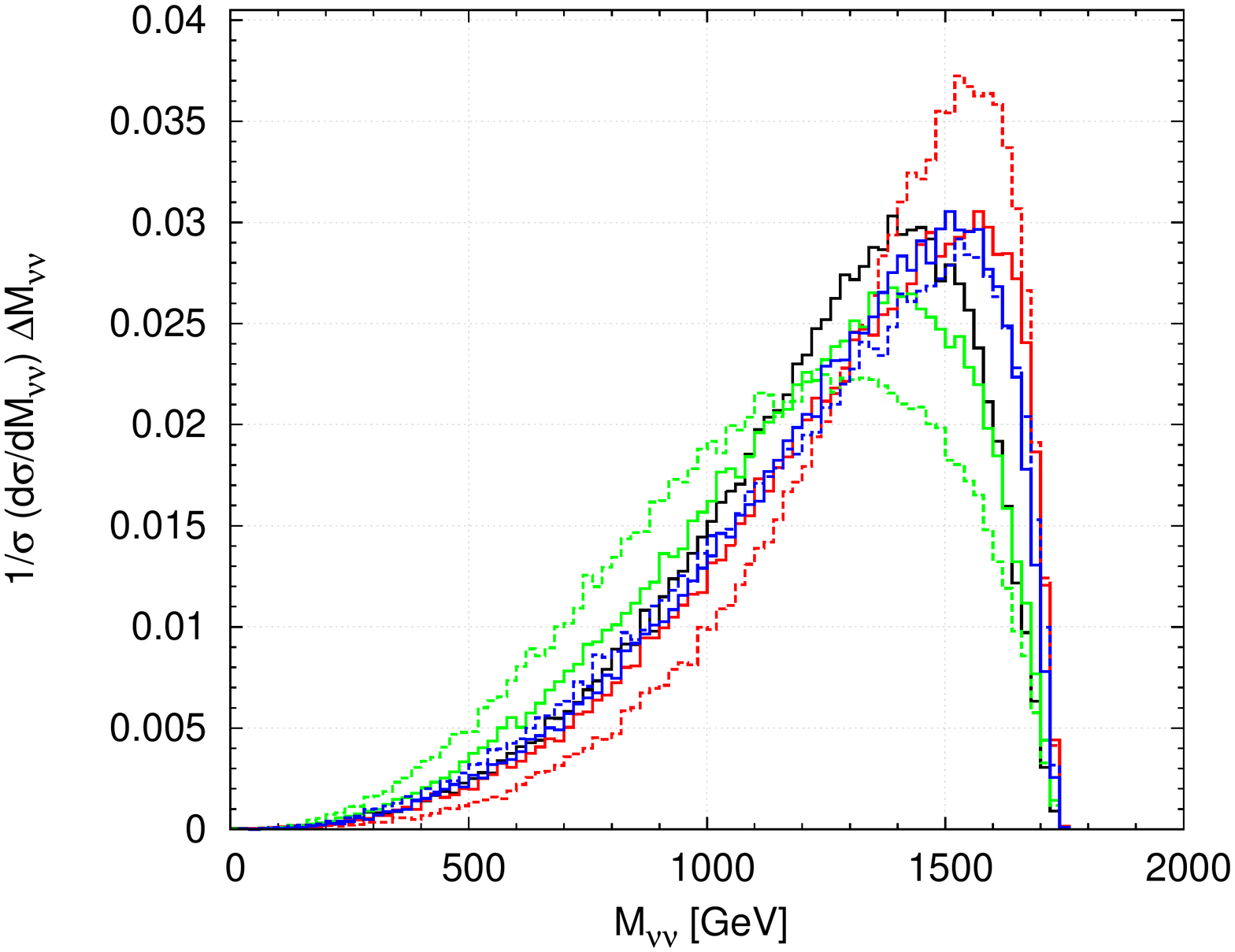} &
\hspace{-8mm}
\vspace{-5mm}
\includegraphics[angle=0,width=80mm]{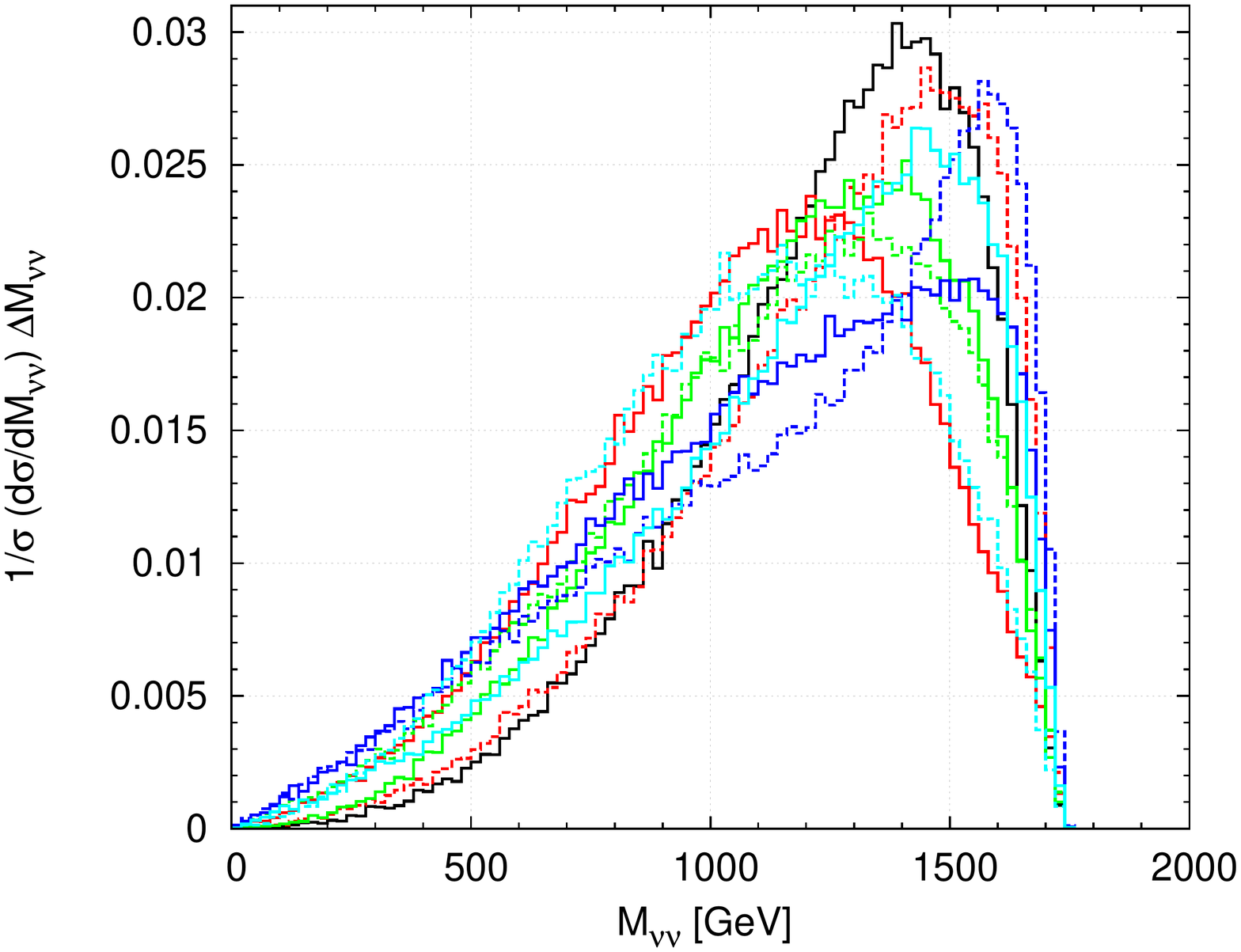}\\
\hspace{-8mm}
\vspace{-5mm}
\includegraphics[angle=0,width=80mm]{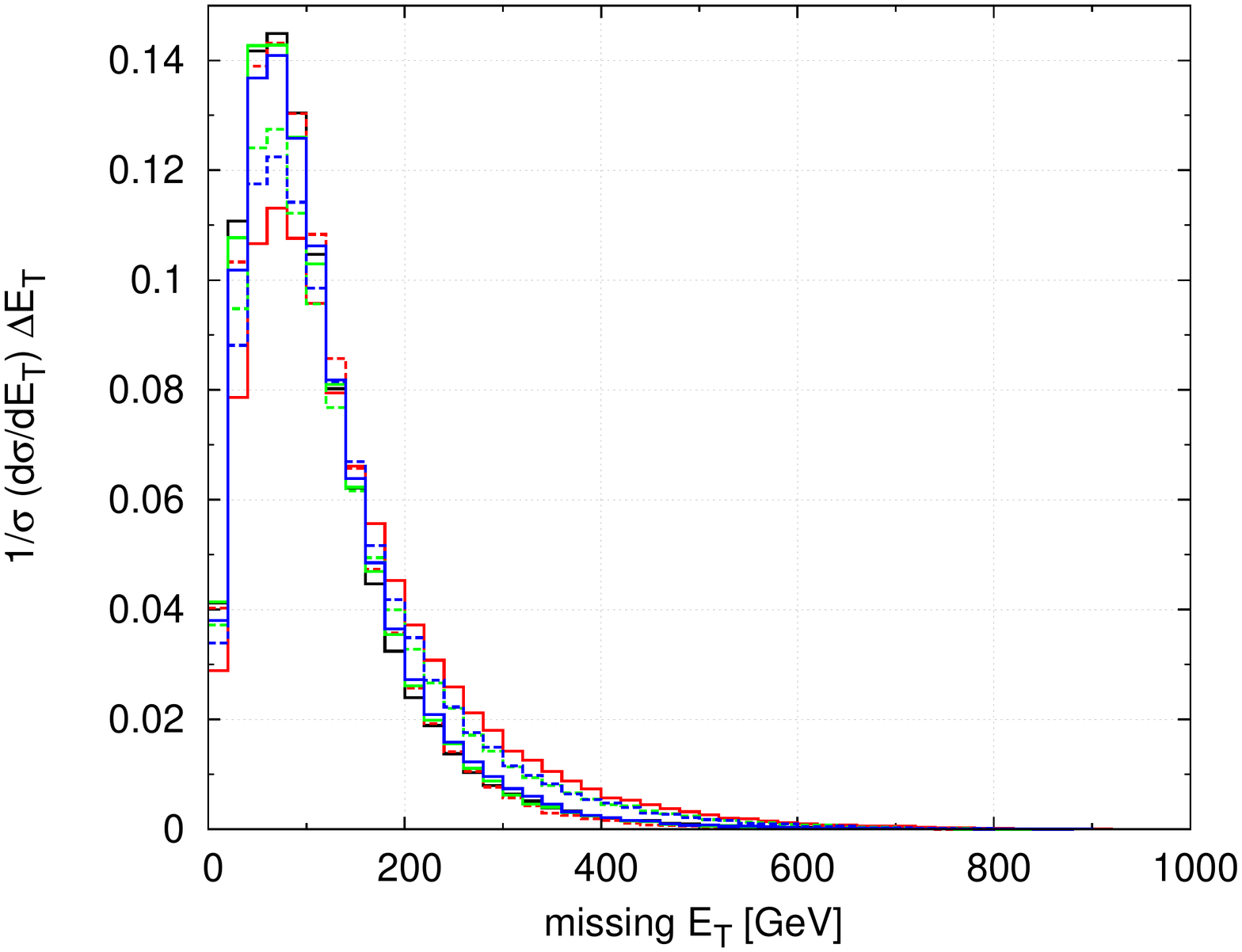} &
\hspace{-5mm}
\vspace{-5mm}
\includegraphics[angle=0,width=80mm]{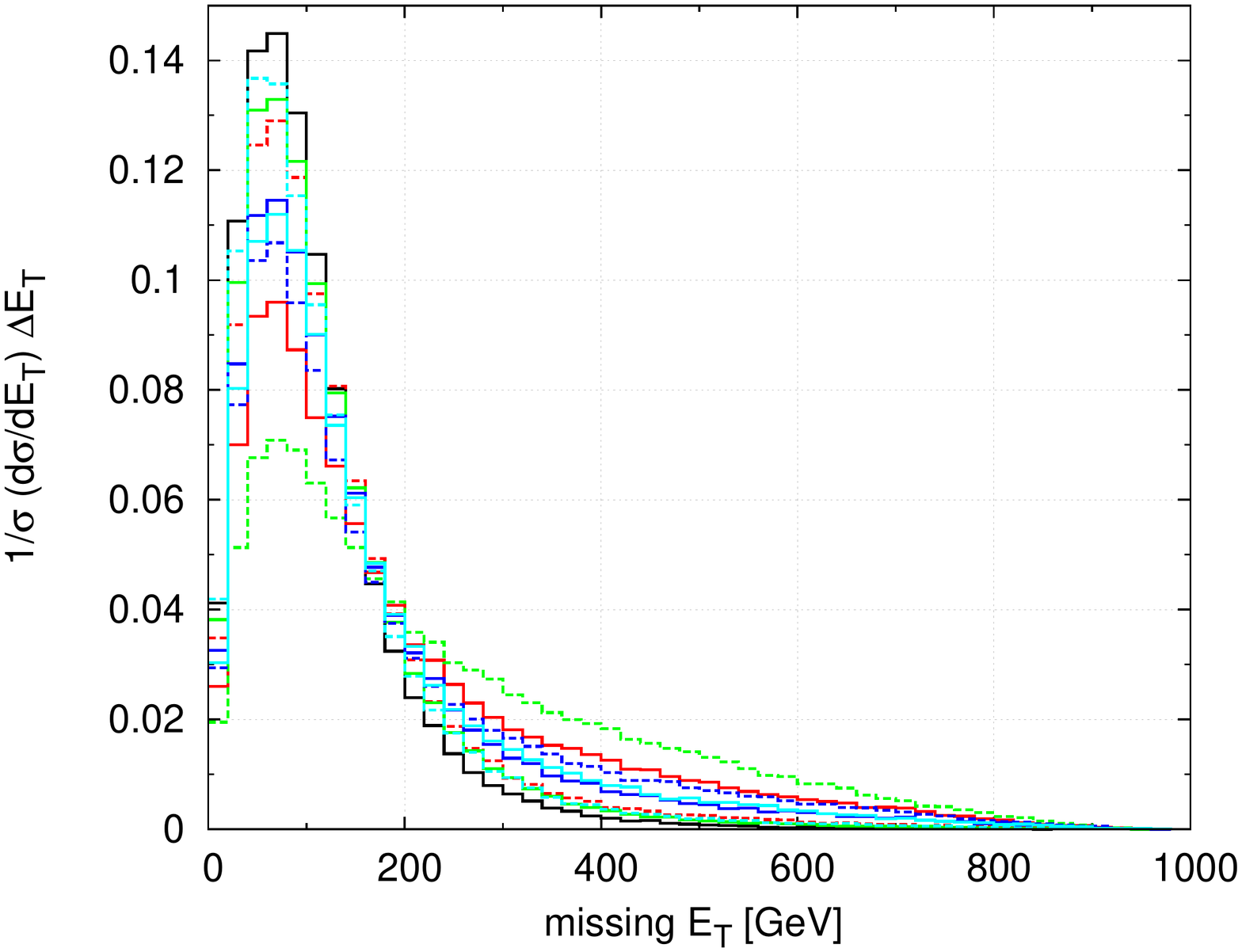}
\end{tabular}
\caption{Kinematic distributions with  the anmalous coupling values as in the inset, illustrating how the presence of $c_W$ (first column), and  $c_{HW}$ and $c_{HB}$ (second column) affect the influence of $c_6$ and $c_H$. A centre of mass energy of  $800$ GeV is assumed.}
\label{fig:ctHH_Mhh_Mnn}
\end{figure}

Moving on to the kinematic distributions, we shall present the distributions of the opening angle between the two Higgs bosons is presented in Fig.\ref{fig:ctHH_Mhh_Mnn} (first row).  The effect of $c_W$, and $c_{HW}$ and $c_{HB}$ are presented separately in the first column and the second column, respectivley. In both cases, the case with only $c_6$ and $c_H$ considered to be non-vanishing, and the SM case are presented for comparison. The dependence of the gauge-Higgs coupling on the sensitivity of  $HHH$ coupling is clear from the plots. The $HH$ inviariant mass as well as the missing invariant mass distributions also indicate a similar dependence, as presented in Fig.\ref{fig:ctHH_Mhh_Mnn} (second row) and (third row). On the other hand, the missing transverse energy distribution does not show much influence of the Higgs-gauge couplings ont he sensitivity of $c_6$ and $c_H$. 

\section{Summary and Conclusions}\label{sec:summary}

The recent discovery of the Higgs boson at LHC has established the Higgs mechanism as the way to have electroweak symmetry breaking, thus generating masses to all the particles. While the mass of the particle is more or less preceisely measured, details like the strenghts of its self interactions, its couplings with other particles like the gauge bosons, etc. need to be know precisely to understand and pinpoint the exact mechanism of electroweak symmetry breaking. Precise knowledge of the trilinear Higgs self-coupling, which is typically probed directly through processes involving two Higgs production, play a vital role in reconstructing the Higgs potential. Typically, such processes also involve other couplings from the Higgs sector, like the Higgs-gauge boson couplings. We consider the $ZHH$ and $\nu \bar\nu HH$ productions at ILC to understant the influence of the $ZZH$ and $ZZHH$ couplings, in the first process, and $WWH$ and $WWHH$ couplings, on the second process, on the sensitivity of $HHH$ coupling on this process. Single and two parameter limits on the $c_6$ and $c_H$ couplings, which are related to the $HHH$ couplings, are considered in the case of an ILC with $\sqrt{s}=800$ GeV and integrated luminosity of 1000 fb$^{-1}$, to see how the other parameters, $c_W$, $c_{HW}$ and $c_{HB}$ influence the limits. It is seen that these latter parameters have significant influence of the reach of $c_6$ and $c_H$, indicating that prior, and somewhat precise knowedge of the Higgs-gauge coupling is necessary to draw any conclusion on the influence of trilinear couplings on the process considered. The kinematic distributions also indicate a strong influence of Higgs-gauge couplings, showing that, in the presence of very moderate Higgs-gauge couplings, it is difficult to extract reliable information regarding $c_6$ and $c_H$. A similar story is unfolded by considerations of $e^+e^-\rightarrow \nu \bar \nu HH$, where the influence of $WWH$ and $WWHH$ on the sensitivity of the trilinear Higgs self-coupling is explored. Concluding, one may need to rely on knowledge of the  Higgs gauge couplings from elsewhere, or consider clever observables eliminating or subduing their effects, in order to extract meaningful information regarding the trilinear Higgs couplings.


\begin{thebibliography}{99}

\bibitem{cms}
  S.~Chatrchyan {\it et al.}  [CMS Collaboration],
  Phys.\ Lett.\ B {\bf 716}, 30 (2012)
  [arXiv:1207.7235 [hep-ex]].

\bibitem{atlas}
  G.~Aad {\it et al.}  [ATLAS Collaboration],
  Phys.\ Lett.\ B {\bf 716}, 1 (2012)
  [arXiv:1207.7214 [hep-ex]].
\bibitem{Moriond1}
The Atlas Collaboration, ATLAS-CONF-2013-029,\,
http://cds.cern.ch/record/1527124/files/ATLAS-CONF-2013-029.pdf

\bibitem{Moriond2}
The Atlas Collaboration, ATLAS-CONF-2013-013,\,
http://cds.cern.ch/record/1523699/files/ATLAS-CONF-2013-013.pdf


\bibitem{Moriond3}
The Atlas Collaboration, ATLAS-CONF-2013-031,\,
http://cds.cern.ch/record/1527127/files/ATLAS-CONF-2013-031.pdf

\bibitem{Moriond4}
The CMS Collaboration, HIG-13-002-pas,\,
http://cds.cern.ch/record/1523767/files/HIG-13-002-pas.pdf

\bibitem{Moriond5}
The CMS Collaboration, HIG-13-003-pas,\,
http://cds.cern.ch/record/1523673/files/HIG-13-003-pas.pdf

\bibitem{Aad:2013wqa}
  G.~Aad {\it et al.}  [ATLAS Collaboration],
  Phys.\ Lett.\ B {\bf 726} (2013) 88
  [arXiv:1307.1427 [hep-ex]].

\bibitem{Chatrchyan:2012ufa}
  S.~Chatrchyan {\it et al.}  [CMS Collaboration],
  Phys.\ Lett.\ B {\bf 716} (2012) 30
  [arXiv:1207.7235 [hep-ex]].

\bibitem{Chatrchyan:2013lba}
  S.~Chatrchyan {\it et al.}  [CMS Collaboration],
  JHEP {\bf 1306} (2013) 081
  [arXiv:1303.4571 [hep-ex]].
\bibitem{Aad:2012tfa}
  G.~Aad {\it et al.}  [ATLAS Collaboration],
  Phys.\ Lett.\ B {\bf 716} (2012) 1
  [arXiv:1207.7214 [hep-ex]].

\bibitem{ILC1}
  J.~Brau, (Ed.) {\it et al.}  [ILC Collaboration],
  arXiv:0712.1950 [physics.acc-ph].

\bibitem{ILC2}
  G.~Aarons {\it et al.}  [ILC Collaboration],
  arXiv:0709.1893 [hep-ph].

\bibitem{polarizationreview}
  G.~Moortgat-Pick, T.~Abe, G.~Alexander, B.~Ananthanarayan,
A.~A.~Babich, V.~Bharadwaj, D.~Barber and A.~Bartl {\it et al.},
  Phys.\ Rept.\  {\bf 460}, 131 (2008)
  [hep-ph/0507011].
\bibitem{Ananthanarayan:2014eea}
  B.~Ananthanarayan, S.~K.~Garg, C.~S.~Kim, J.~Lahiri and P.~Poulose,
  Phys.\ Rev.\ D {\bf 90} (2014) 014016
  [arXiv:1405.6465 [hep-ph]].
\bibitem{Ananthanarayan:2013cia}
  B.~Ananthanarayan, S.~K.~Garg, J.~Lahiri and P.~Poulose,
  Phys.\ Rev.\ D {\bf 87} (2013) 11,  114002
  [arXiv:1304.4414 [hep-ph]].
\bibitem{Muhlleitner:2012jy}
  M.~Muhlleitner, R.~M.~Godbole, C.~Hangst, S.~D.~Rindani and P.~Sharma,
  Frascati Phys.\ Ser.\  {\bf 54} (2012) 188.

\bibitem{Godbole:2011hw}
  R.~M.~Godbole, C.~Hangst, M.~Muhlleitner, S.~D.~Rindani and P.~Sharma,
  Eur.\ Phys.\ J.\ C {\bf 71} (2011) 1681
  [arXiv:1103.5404 [hep-ph]].
\bibitem{Weinberg:1978kz}
  S.~Weinberg,
  Physica A {\bf 96} (1979) 327.
\bibitem{Weinberg:1980wa}
  S.~Weinberg,
  Phys.\ Lett.\ B {\bf 91} (1980) 51.
\bibitem{Georgi:1994qn}
  H.~Georgi,
  Ann.\ Rev.\ Nucl.\ Part.\ Sci.\  {\bf 43} (1993) 209.
\bibitem{Buchmuller:1985jz}
  W.~Buchmuller and D.~Wyler,
  Nucl.\ Phys.\ B {\bf 268} (1986) 621.
\bibitem{Hagiwara:1993ck}
  K.~Hagiwara, S.~Ishihara, R.~Szalapski and D.~Zeppenfeld,
  Phys.\ Rev.\ D {\bf 48} (1993) 2182.
\bibitem{Hagiwara:1993qt}
  K.~Hagiwara, R.~Szalapski and D.~Zeppenfeld,
  Phys.\ Lett.\ B {\bf 318} (1993) 155
  [hep-ph/9308347].
\bibitem{Alam:1997nk}
  S.~Alam, S.~Dawson and R.~Szalapski,
  Phys.\ Rev.\ D {\bf 57} (1998) 1577
  [hep-ph/9706542].
\bibitem{genuined6} 
  V.~Barger, T.~Han, P.~Langacker, B.~McElrath and P.~Zerwas,
  Phys.\ Rev.\ D {\bf 67}, 115001 (2003)
  [hep-ph/0301097].
\bibitem{Giudice:2007fh}
  G.~F.~Giudice, C.~Grojean, A.~Pomarol and R.~Rattazzi,
  JHEP {\bf 0706} (2007) 045
  [hep-ph/0703164].
\bibitem{Contino2010a}
R. Contino, C. Grojean, M. Moretti, F. Piccinini and
R. Rattazzi, JHEP 1005 (2010) 089 [arXiv:1002.1011 [hep-ph]]; 
\bibitem{Contino2010b}
R. Contino, arXiv:1005.4269 [hep-ph]; 
\bibitem{Grober2011}
R. Grober and M. Muhlleitner, JHEP 1106 (2011) 020 [arXiv:1012.1562 [hep-ph]].
\bibitem{Grzadkowski:2010es}
  B.~Grzadkowski, M.~Iskrzynski, M.~Misiak and J.~Rosiek,
  JHEP {\bf 1010} (2010) 085
  [arXiv:1008.4884 [hep-ph]].
\bibitem{GutierrezRodriguez:2011gi}
  A.~Gutierrez-Rodriguez, J.~Peressutti and O.~A.~Sampayo,
  J.\ Phys.\ G {\bf 38} (2011) 095002
  [arXiv:1107.0245 [hep-ph]].

\bibitem{GutierrezRodriguez:2009uz}
  A.~Gutierrez-Rodriguez, M.~A.~Hernandez-Ruiz and O.~A.~Sampayo,
Energies,''
  Int.\ J.\ Mod.\ Phys.\ A {\bf 24} (2009) 5299
  [arXiv:0903.1383 [hep-ph]].

\bibitem{GutierrezRodriguez:2005fe}
  A.~Gutierrez-Rodriguez, M.~A.~Hernandez-Ruiz and O.~A.~Sampayo,
  Mod.\ Phys.\ Lett.\ A {\bf 20} (2005) 2629
  [hep-ph/0504266].

\bibitem{Rindani:2010pi}
  S.~D.~Rindani and P.~Sharma,
  Phys.\ Lett.\ B {\bf 693} (2010) 134
  [arXiv:1001.4931 [hep-ph]].

\bibitem{Rindani:2009pb}
  S.~D.~Rindani and P.~Sharma,
  Phys.\ Rev.\ D {\bf 79} (2009) 075007
  [arXiv:0901.2821 [hep-ph]].

  
\bibitem{Baak:2012kk}
  M.~Baak, M.~Goebel, J.~Haller, A.~Hoecker, D.~Kennedy, R.~Kogler, K.~Moenig and M.~Schott {\it et al.},
  Eur.\ Phys.\ J.\ C {\bf 72} (2012) 2205
  [arXiv:1209.2716 [hep-ph]].
\bibitem{Einhorn:2013kja}
  M.~B.~Einhorn and J.~Wudka,
  Nucl.\ Phys.\ B {\bf 876} (2013) 556
  [arXiv:1307.0478 [hep-ph]].
\bibitem{Contino:2013kra}
  R.~Contino, M.~Ghezzi, C.~Grojean, M.~Muhlleitner and M.~Spira,
  JHEP {\bf 1307} (2013) 035
  [arXiv:1303.3876 [hep-ph]].
  
\bibitem{Amar:2014fpa}
  G.~Amar, S.~Banerjee, S.~von Buddenbrock, A.~S.~Cornell, T.~Mandal, B.~Mellado and B.~Mukhopadhyaya,
  arXiv:1405.3957 [hep-ph].
\bibitem{Masso:2014xra}
  E.~Masso,
  arXiv:1406.6376 [hep-ph].

\bibitem{Biekoetter:2014jwa}
  A.~Biekoetter, A.~Knochel, M.~Kraemer, D.~Liu and F.~Riva,
  arXiv:1406.7320 [hep-ph].
\bibitem{Willenbrock:2014bja}
  S.~Willenbrock and C.~Zhang,
  arXiv:1401.0470 [hep-ph].


\bibitem{Bonnet:2011yx}
  F.~Bonnet, M.~B.~Gavela, T.~Ota and W.~Winter,
  Phys.\ Rev.\ D {\bf 85} (2012) 035016
  [arXiv:1105.5140 [hep-ph]].
\bibitem{Corbett:2012dm}
  T.~Corbett, O.~J.~P.~Eboli, J.~Gonzalez-Fraile and M.~C.~Gonzalez-Garcia,
  Phys.\ Rev.\ D {\bf 86} (2012) 075013
  [arXiv:1207.1344 [hep-ph]].
\bibitem{Chang:2013cia}
  W.~-F.~Chang, W.~-P.~Pan and F.~Xu,
  Phys.\ Rev.\ D {\bf 88} (2013) 3,  033004
  [arXiv:1303.7035 [hep-ph]].
\bibitem{Elias-Miro:2013mua}
  J.~Elias-Miro, J.~R.~Espinosa, E.~Masso and A.~Pomarol,
  JHEP {\bf 1311} (2013) 066
  [arXiv:1308.1879 [hep-ph]].
\bibitem{Banerjee:2013apa}
  S.~Banerjee, S.~Mukhopadhyay and B.~Mukhopadhyaya,
  Phys.\ Rev.\ D {\bf 89} (2014) 053010
  [arXiv:1308.4860 [hep-ph]].
\bibitem{Boos:2013mqa}
  E.~Boos, V.~Bunichev, M.~Dubinin and Y.~Kurihara,
  Phys.\ Rev.\ D {\bf 89} (2014) 3,  035001
  [arXiv:1309.5410 [hep-ph]].
 
\bibitem{Masso:2012eq}
  E.~Masso and V.~Sanz,
  Phys.\ Rev.\ D {\bf 87} (2013) 3,  033001
  [arXiv:1211.1320 [hep-ph]].
\bibitem{Han:2004az}
  Z.~Han and W.~Skiba,
  Phys.\ Rev.\ D {\bf 71} (2005) 075009
  [hep-ph/0412166].
\bibitem{Corbett:2012ja}
  T.~Corbett, O.~J.~P.~Eboli, J.~Gonzalez-Fraile and M.~C.~Gonzalez-Garcia,
  Phys.\ Rev.\ D {\bf 87} (2013) 015022
  [arXiv:1211.4580 [hep-ph]].
\bibitem{Dumont:2013wma}
  B.~Dumont, S.~Fichet and G.~von Gersdorff,
  JHEP {\bf 1307} (2013) 065
  [arXiv:1304.3369 [hep-ph]].
\bibitem{Pomarol:2013zra}
  A.~Pomarol and F.~Riva,
  JHEP {\bf 1401} (2014) 151
  [arXiv:1308.2803 [hep-ph]].
  \bibitem{Ellis:2014dva}
  J.~Ellis, V.~Sanz and T.~You,
  arXiv:1404.3667 [hep-ph].
\bibitem{Belusca-Maito:2014dpa}
  H.~Belusca-Maito,
  arXiv:1404.5343 [hep-ph].
\bibitem{Gupta:2014rxa}
  R.~S.~Gupta, A.~Pomarol and F.~Riva,
  arXiv:1405.0181 [hep-ph].

\bibitem{Aad:2013wqa}
  G.~Aad {\it et al.}  [ATLAS Collaboration],
  Phys.\ Lett.\ B {\bf 726} (2013) 88
  [arXiv:1307.1427 [hep-ex]].
\bibitem{Teyssier:2014hta}
  D.~Teyssier [ATLAS and CMS Collaborations],
  arXiv:1404.7311 [hep-ex].
Mebane:2013zga, Mebane:2013zga
\bibitem{De Rujula:1991se}
  A.~De Rujula, M.~B.~Gavela, P.~Hernandez and E.~Masso,
  Nucl.\ Phys.\ B {\bf 384} (1992) 3.
\bibitem{1}
  A.~Gutierrez-Rodriguez, M.~A.~Hernandez-Ruiz, O.~A.~Sampayo, A.~Chubykalo and A.~Espinoza-Garrido,
  J.\ Phys.\ Soc.\ Jap.\  {\bf 77} (2008) 094101
  [arXiv:0807.0663 [hep-ph]].
\bibitem{2}
  Y.~Takubo,
  arXiv:0907.0524 [hep-ph].
\bibitem{3}
  J.~Tian, K.~Fujii and Y.~Gao,
  arXiv:1008.0921 [hep-ex].
\bibitem{4}
  M.~Battaglia, E.~Boos and W.~M.~Yao,
  eConf C {\bf 010630} (2001) E3016
  [hep-ph/0111276].
\bibitem{5}
  V.~Barger, T.~Han, P.~Langacker, B.~McElrath and P.~Zerwas,
  Phys.\ Rev.\ D {\bf 67} (2003) 115001
  [hep-ph/0301097].
\bibitem{6}
  R.~Killick, K.~Kumar and H.~E.~Logan,
  Phys.\ Rev.\ D {\bf 88} (2013) 033015
  [arXiv:1305.7236 [hep-ph]].
\bibitem{7}
  A.~Djouadi, W.~Kilian, M.~Muhlleitner and P.~M.~Zerwas,
  Eur.\ Phys.\ J.\ C {\bf 10} (1999) 27
  [hep-ph/9903229].
\bibitem{8}
  H.~Baer, T.~Barklow, K.~Fujii, Y.~Gao, A.~Hoang, S.~Kanemura, J.~List and H.~E.~Logan {\it et al.},
  arXiv:1306.6352 [hep-ph].
\bibitem{9}
  C.~Castanier, P.~Gay, P.~Lutz and J.~Orloff,
  In *2nd ECFA/DESY Study 1998-2001* 1362-1372
  [hep-ex/0101028].
\bibitem{10}K. Fujii, talk given at the Higgs Snowmass Work-
shop, Princeton, New Jersey, USA, Jan. 14-15, 2013, slides available from http://physics.princeton.edu/ indico/conferenceDisplay.py?confId=127.

\bibitem{madgraph}
  J.~Alwall, M.~Herquet, F.~Maltoni, O.~Mattelaer and T.~Stelzer,
  JHEP {\bf 1106} (2011) 128
  [arXiv:1106.0522 [hep-ph]].
\bibitem{feynrules} FeynRules:http://feynrules.irmp.ucl.ac.be/wiki/HEL

\bibitem{HEL}
  A.~Alloul, B.~Fuks and V.~Sanz,
  arXiv:1310.5150 [hep-ph].


\end{thebibliography}
\end{document}